\documentclass[superscriptaddress,preprintnumbers,floatfix,showpacs,showkeys]{revtex4}

\usepackage{epsfig}
\usepackage{subfigure}
\usepackage{amscd}
\usepackage{amsmath}
\usepackage{amssymb}
\usepackage{color}

\definecolor{darkred}{rgb}{.7,0,0}

\definecolor{darkgreen}{rgb}{0,0.7,0}

\definecolor{darkblue}{rgb}{0,0,0.7}

\newlength{\arrayrulewidthOriginal}
\newcommand{\Cline}[2]{%
  \noalign{\global\setlength{\arrayrulewidthOriginal}{\arrayrulewidth}}%
  \noalign{\global\setlength{\arrayrulewidth}{#1}}\cline{#2}%
  \noalign{\global\setlength{\arrayrulewidth}{\arrayrulewidthOriginal}}}

\usepackage{multirow}
\begin{document}

\title{The Parameter Houlihan: a solution to high-throughput identifiability indeterminacy for brutally ill-posed problems}

\author{D. J. Albers}
\email{david.albers@ucdenver.edu}
\affiliation{Department of Pediatrics, Division of Informatics,
  University of Colorado, Aurora, CO 80045}
\affiliation{Department of Biomedical Informatics, Columbia
  University, 622 West 168th Street, PH-20, New York, NY 10032}

\author{M Levine}
\email{mlevine@caltech.edu}
\affiliation{Department of computational and mathematical sciences, 1200 E California Blvd M/C 305-16 Pasadena, CA 91125}

\author{L Mamykina}
\email{om2196@cumc.columbia.edu}
\affiliation{Department of Biomedical Informatics, Columbia
  University, 622 West 168th Street, PH-20, New York, NY 10032}

\author{G. Hripcsak}
\email{hripcsak@columbia.edu}
\affiliation{Department of Biomedical Informatics, Columbia
  University, 622 West 168th Street, PH-20, New York, NY 10032}

\begin{abstract}
One way to interject knowledge into clinically impactful forecasting is to use data assimilation, a nonlinear regression that projects data onto a mechanistic physiologic model, instead of a set of functions, such as neural networks. Such regressions have an advantage of being useful with particularly sparse, non-stationary clinical data. However, physiological models are often nonlinear and can have many parameters, leading to potential problems with parameter identifiability, or the ability to find a unique set of parameters that minimize forecasting error. The identifiability problems can be minimized or eliminated by reducing the number of parameters estimated, but reducing the number of estimated parameters also reduces the flexibility of the model and hence increases forecasting error.  We propose a method, the parameter Houlihan, that combines traditional machine learning techniques with data assimilation, to select the right set of model parameters to minimize forecasting error while reducing identifiability problems. The method worked well: the data assimilation-based glucose forecasts and estimates for our cohort using the Houlihan-selected parameter sets generally also minimize forecasting errors compared to other parameter selection methods such as by-hand parameter selection. Nevertheless, the forecast with the lowest forecast error does not always accurately represent physiology, but further advancements of the algorithm provide a path for improving physiologic fidelity as well. Our hope is that this methodology represents a first step toward combining machine learning with data assimilation and provides a lower-threshold entry point for using data assimilation with clinical data by helping select the right parameters to estimate.
\end{abstract}


\maketitle

% \linenumbers

\section{Introduction}

We want to use data and our understanding of the world to better manage health --- we want evidence and understanding to guide clinical and personal health-related decisions. Of course at a high level this is generally what medicine is about: interventions are undertaken only when they are understood or predicted to improve an individual's health. However, traditionally this prediction is done in a non-personalized manner, meaning that interventions treat the "mean" person or patient. Personalized and precision medicine were conceptualized to relax this constraint by tailoring an intervention to a person. While genetics offers a path to personalizing treatment, we can also use data science machinery together with personal (\cite{da_glucose_forecast_t2d}) and population-scale data to better personalize treatment (\cite{pop_phys,pmlr-v56-Xu16,dyn_pheno}). Specifically here, we want to leverage our knowledge encapsulated in mechanistic physiologic models and combine it with free living or clinical data to allow this knowledge and data to be used to make decisions related to health. In this context, computational problems related to personalized medicine can be broken into two broad categories: \emph{forecasting}, where we make quantitative predictions about a patient's future state that can be used by clinicians and patients to take corrective action, and \emph{phenotyping} (\cite{high_fide_pheno,emerge3,phenome_model,anchor_phenome_jamia,google_deep_pheno}), where we identify properties of macroscopic observables that can be used to classify patients into subgroups that can give clinicians and researchers actionable insight into commonly occurring treatment outcomes and biological phenomena.

The idea of using mechanistic models and data assimilation in biomedicine or healthcare is old, but what is new is attempting to integrate models with variable complexity with sparse, noisy  free-living and clinically collected data.  Many mathematical biology models were designed to have variable degrees of biological fidelity, fidelity that we do not necessarily want to eliminate or reduce, but fidelity that we generally need to constrain in the usual case where we cannot estimate all the parameters because of data limitations that always exist in practice.  This problem poses a significant barrier to using data assimilation---enough of a barrier that often data assimilation is not even attempted because the models, given data are hopelessly poorly resolved. This paper poses a machine learning solution to this problem---by using machine learning to identify and rank-order which model parameters are the most necessary to estimate.

Returning to the more practical contexts of phenotyping and forecasting, both applications impose particular demands on certain aspects of computational machinery used to model data. The properties we focus on here are the selection of the model parameters to estimate and the ensuing \emph{identifiability} of a model, or ability to uniquely solve for parameters that yield optimal solutions (\cite{physio_uncertainty_book,system_id_ljun,linear_id_nonlinear_setting}). Our goal is to strike a balance between identifiability and model fidelity in situations where a model is not fully identifiable if all or even sometimes when any model parameters are estimated, given the available data. The method we develop here can facilitate both forecasting and phenotyping studies, and we evaluate this method in the context of modeling glucose dynamics using mechanistic models, machine learning and data assimilation.

% One way to get at this is to use a regression.
%Supervised methods that attempt to perform personalized forecasting or phenotyping can broadly be conceptualized as regressions between patient information and patient outcomes.

The Houlihan, or the Houlihan throw, is a lasso throw used for roping livestock, e.g., a horse.  It is used often under difficult circumstances such as picking out, from a substantial distance, a single horse from among a crowd of horses standing close together. It is a particularly flexible technique that can be used in a variety of circumstances. In this spirit we intuitively define the Houlihan method(s) as a collection of methods that use for selecting the most productive model parameters to estimate; specifically, the collection of methods uses machine learning techniques applied to simulated model output under parameter variation subject to a set of features, e.g., the mean of a state.

\section{Background}

The larger biomedical context of this work is the application of data science machinery used to personalize forecasts and phenotypes via a broadly defined regression. While there are many linear versions of regression that have been successfully applied to healthcare data (\cite{amia_matt_george_granger_like,full_lagged_regression_matt,george_lagged_correlation_jamia,jamia_hcpmodel13}), here we focus on a specific type of nonlinear regression---data assimilation---in an effort to take advantage of potentially important nonlinearities present in most biological systems. Nonlinear regression approaches such as deep learning and related methods (\cite{deep_survival_I,deep_survival_II,google_deep_pheno,lasko_plos}) have seen some success in a number of biomedical applications thanks to their ability to approximate arbitrary, non-linear functions. While the flexibility of universal approximator approaches (\cite{hor,hor2}) is particularly useful when little is known about the system and data are plentiful, this approach does not always work well when data are sparse and non-stationary, leading to problems such as poor generalization to new or unobserved individuals, problems with quickly changing health conditions, and difficulties with fast, accurate prediction with very few, e.g., $20$, data points. Unfortunately, many health data and healthcare situations fit one or more of these data pathologies (\cite{jamia_phys_ehr,measurement_dynamics_rimma}).

In order to exploit the complex yet rich quantity of available health data, it is natural to consider ways of constraining the search space for machine learning methods. One way to do this is to constrain the model search space in accordance with as much expert knowledge as possible. To achieve this here we turn to mechanistic models developed by mathematical biologists and physiologists \cite{keenerII}, which are typically formulated as dynamical systems (\cite{brin_ds_book,guck,arrowsmithandplace}), e.g., $x_{t+1} = f(x_t,\theta)$, or differential equations (\cite{arnoldode,arnoldgeo}), e.g., $\frac{dx}{dt} = f(x,\theta,t)$, where $x$ are the time-varying states of the system and $\theta$ are the physiologic parameters that govern the process. For example, in the case of phenotyping type 2 diabetes one way of constraining the search space of a regression is to regress the data onto a nonlinear physiologic model \cite{da_glucose_forecast_t2d,daJAMIA} instead of regressing the data onto a universal approximator \cite{hor,hor2} function space such as neural networks. The way this is done is using data assimilation.

Data assimilation (DA) is a collection of methods (\cite{filtering_jaz,data_assimilation_I,stuart_da,french_DA_asch,da_sebastian,bayesian_estimation_tracking,beyond_kalman,baysian_signal_processing,DA_aos_evensen,evensen_enkf_early}) concerned with performing the types of non-linear regressions we describe for dynamical systems, and centers itself around forecasting and inferring mechanistic states under available observations; it solves both forward and inverse problems (\cite{andrew_inverse_problems,clermont_inverse_problems,inverse_book_banks}). There have been many successful applications of mechanistic modeling and data assimilation in biomedicine (\cite{daJAMIA,hirata_prostate_model,hirata_prostate_cancer_chaos,schiff_neuro_control_theory,vanja_sir,SIAM_identifiability_bio_modeling,closed_loop_glucose_control,artificial_pancreas_I,artificial_pancreas_II,in_silico_glucose_model_fda,rubella_model,defib_nejm,cc_art_beta,closed_loop_glucose_control_ICU,corbelli_review,artificial_beta_cell,leon_court,dave_leon_3,christini_cardio_alternans,mackey_glass_equations,clermont_inverse_problems,parker_ICU_1,parker_doyle,pharmaco_2,pharmaco_1,pharmaco_3,pharmaco_4,albers_IEEE,selgrade_female_endo_data,bruce_sleep,chase_icu_2,chase_icu_1,da_glucose_forecast_t2d,online_offline_DA}). However, mechanistic models that are typically developed in biological laboratory settings are often not designed to interface with health data collected in the process of delivering care or in free-living situations---in particular, the physiologic models often model macroscopic states that \emph{are} observable from routinely collected data but are governed by a composition of \emph{unobservable} mechanisms. While these models capture the dynamics we are interested in and constrain the regression to a smaller class of functions, their high-fidelity creates issues of identifiability and ill-posedness, problems for which this paper develops a practical, machine-learning-based work-around.

% While data assimilation with a mechanistic model significantly constrains the class of functions permitted to represent the data, it can still suffer from issues of identifiability.

% Parameter identifiability of neural nets and other function approximators is not typically a concern---the goal is simply to find some parameter region that provides an adequate approximation. This is acceptable practice because these models are very easy to optimize using stochastic gradient descent. However, non-linear dynamical systems

% and in particular, data assimilation (DA), or nonlinear regressions where the data are projected onto a mechanistic model instead of a more general function space such as neural networks.  We are focused on DA with an eye on problems where the data are sparse, non-stationary, and may not generalize well between individuals---situations where nonlinear regressions can be problematic but where DA can have certain advantages due to the use of models constructed with human insight.  The advantages come with disadvantages, of course --- the models do not perfectly represent nature, can have non-unique solutions in parameter space (i.e., fail to be identifiable), etc., but at the core, the problem lies in how to represent nature in a way that can be exploited to make reliable predictions
% with small amounts of data.

To understand how identifiability works for these machines, consider a trivial case of identifiability for the model $\frac{dx}{dt} = abx$. If we assume that $a$ and $b$ are unknown parameters, they cannot both be identifiable without another equation that could uniquely determine one of them. This topic and the the associated methods for handling this situation are too old and wide ranging to give complete background (\cite{system_id_ljun,physio_uncertainty_book,physio_bayesian_identifiability,linear_id_nonlinear_setting}). We can, however, give a broad sketch of how identifiability has been traditionally approached. Identifiability analysis generally follows one of three pathways: \emph{analytical methods}, e.g., showing algebraically that all parameters can be uniquely solved for (\cite{cecelia_ogtt_ident,marissa_ident,marissa_2});  \emph{numerical methods} (\cite{linear_id_nonlinear_setting,physio_uncertainty_book,marissa_3}); and \emph{heuristic, knowledge-driven} sensitivity analyses where certain parameters are chosen based on computational experiments or knowledge of the system. In many complex, non-linear mechanistic physiologic models algebraic methods and linear computational methods are not tractable or applicable.  In these situations nonlinear methods can be applied, but nonlinear methods usually have to be constructed for a particular situation (\cite{structural_id_hard}), and, much like nonlinear optimization, generally do not have clearcut or simple resolutions (\cite{structural_id_hard,physio_bayesian_identifiability}). \emph{These problems pose a significant roadblock to parameter inference in the context of DA.} Nevertheless, there exist methods for working to remain within a traditional identifiability framework, e.g., \cite{physio_bayesian_identifiability} uses Bayesian inference to determine when parameters can be made identifiable.

The usual way of addressing these issues focuses on making sure the model is identifiable or finding ways of making it more identifiable (\cite{physio_bayesian_identifiability,linear_id_nonlinear_setting,physio_uncertainty_book}). This work is often performed using substantial intuition about the important features encoded in the model, and parameterizing and grouping sub-processes. However, this creates silos of expertise and prevents wide-spread dissemination and evaluation of mechanistic models in  potential application domains. Therefore, to progress toward understanding complex physiology via model refinement and selection, and to provide solutions in clinical situations that come with constraints of time-sensitive solutions, we must find a robust way of coping with brutally ill-posed problems and accept certain impurities and inaccuracies.
% In other words, to facilitate wider and more flexible usage, we need to develop a more high-throughput approach.

Here, we develop and evaluate a method for rank-ordering mechanistic parameters based on their "influence" on important dynamical features, in order to improve forecast accuracy and help determine which models most faithfully represent a given system. This provides a starting point from which to estimate parameters, prune the model, etc., that can be automated.

\section{Conceptual construction of the Houlihan approach}

\subsection{Conventional operational use of data assimilation with ill-posed problems}

The standard method of applying data assimilation (DA) or control in generic situations follows roughly the following steps: (i) select a model, (ii)  work out identifiability, (iii) select a filter or inference method, (iv) find an optimal solution for states and parameters. This requires very careful experimental constructions, generally dense data streams, can be expensive, and requires relatively simple models, all situations that lie outside of what is possible in applications and even many basic science settings. The approach for applying DA in \emph{operational}, complex, high-dimensional settings where accurate \emph{real-time} forecasts are imperative is to: (i) select or develop a model, (ii) tune and fix parameters offline, often by hand or using a combination of by-hand and numeric tuning that allows the model to reconstruct or forecast states within some tolerance, (iii) select an inference scheme, and (iv)  estimate states only and make a forecast. This is a tried and true method and is used in situations such as weather and climate forecasting (\cite{tom_enkf_review,DA_aos_evensen,evensen_enkf_early}). Neither of these approaches apply to biomedical situations that, by comparison, have a different set of constraints and problems, including: (i) the models are smaller, so they can be simulated faster and estimated faster, allowing for potentially many models to be used simultaneously; (ii) there are less data relative to the number of \emph{unknown} parameters, so while parameter estimation is necessary \cite{da_glucose_forecast_t2d} not all parameters can be estimated; (iii) models are not generated from first principles and their application to given individuals is potentially highly variable necessitating the use and potentially the averaging of many models; and (iv) tuning would have to be done for millions of people frequently, e.g., every patient in every ICU potentially every day, a process that is not likely to be practically possible. Because of these reasons, choosing which parameters to estimate is a significant barrier to the adoption and use of DA in biomedical situations.

% and biological modeling \cite{lungguy,hines,phys_book_parm_est}.

\subsection{Houlihan approach to ill-posedness}

Here we are operating under a different situation from the more canonical DA application setting, one more heavily constrained by imperfections of data that will never disappear because the data are collected in the process of managing health instead of data collected in a controlled manner explicitly for the DA. In particular we assume: \emph{(a)} we do not know the right model but we have some models we can try, \emph{(b)} we do not know whether a given model is identifiable and that we do not have enough data to estimate all model parameters well anyway but that we have enough data to estimate at least one parameter, \emph{(c)} for a given model, we don't know what parameters are the most useful to estimate, given that we cannot estimate all of them. Given this situation we develop a method for rank ordering which parameters to estimate, subject to features we want to capture, when we have no idea how to choose which parameters to estimate or when we must choose parameters in a more high-throughput setting where we are using many models at once.

This solution involves stacking machine learning on top of DA: machine learning is applied to \emph{simulated} model output to select the important parameters to estimate to best synchronize the model with the data, and then we use DA restricted to estimation of the parameters chosen by machine learning. In this way, the method will scale to a high-throughput setting and can be applied to many different models with high dimensional parameter spaces more easily. And while we know that this method may not lead to a unique solution in function, parameter, or initial condition space, the set of solutions will be reduced to a workable set of solutions that allow forward progress to be made.

Conceptually, we are proposing to: \emph{(i)} assume a model, \emph{(ii)} simulate the model under discrete parameter variation creating a grid in parameter space for which at every point we have simulated data from the model (i.e., the instance of one attractor of the model for a given set of initial conditions at the parameter grid point), \emph{(iii)} select features, e.g., the mean, of the attractors that are important for estimating the physiologic system, and then \emph{(iv)} use a machine learning algorithm to identify the parameters that have the greatest impact on the features. While for the authors the geometric intuition of this method originates from bifurcation theory---we will discuss this in a later section \cite{kuzbook}---one useful way to think about the problem is in the inverse problems context. \emph{As was the case for the bifurcation theory context, this discussion is allegorical; we are not proposing a formal inverse problems regularization framework.} From a high level, given data, $Y$, and a model, $\mathcal{F}$ with a state space $x$, the task is to find a \emph{set} of parameter values, $\Theta_i$, of which there may be many if the system is not identifiable, that minimize:
\begin{equation}
\label{eq:min}
|| Y - \mathcal{F}(x, \theta)||^p_{Y}
\end{equation}
for some $p$, $p=2$ being the commonly applied least squares minimization. The core of the identifiability issue is that, for complex models, and especially given sparse data, there may be many sets of parameters $\Theta_i$ that minimize the distance between the model and the data. In this case a goal might then be to balance the number of potential minimizing parameter sets, the number of $\Theta_i$'s, against the distance between the model and the data via an optimization algorithm, e.g.,
\begin{equation}
\min_{\Theta} (w_1( || Y - \mathcal{F}(x, \theta)||^p_{Y} )+ w_2( \# \{  \Theta_i \} ))
\end{equation}
where the $w_i$'s are continuous functions. This framework, a formal regularization methodology, has many advantages, but can induce many complexities that increase rather than decrease the barrier to using data assimilation in more data-poor environments. Moreover, this relatively complex methodology may not be applicable in more high-throughput situations where, e.g., many models are used in a model averaging context. Therefore, motivated by the goal of an imperfect but practical solution, we postulate that if we carefully select the right parameters that maximize the parameter subspaces that can be explored relative to a set of desired features, we can often, effectively but imperfectly, solve the optimization problem. Effectively but not rigorously, we are \emph{regularizing a priori}, by selecting and reducing the parameter set to be estimated before we go about estimating the parameters given data. Given the framework above, such a solution may be well handled by a tool from sparse machine learning such as lasso \cite{statistical_learning_sparsity} because it uniquely rank-orders parameters by their predictive power, but it is easy to imagine using other methods. But, it is important to be clear that we are hypothesizing that the parameter subspaces that allow maximal exploration of dynamics relative to a given feature, e.g., the mean, will contain sets of parameters, $\Theta_i$ that also find relatively good minima of Eq. \ref{eq:min}. In our evaluation we will see cases where this hypothesis fails, but we will also see that this hypothesis generally holds true in our data set, and this conclusion is the point of the paper.

In short, here we are assuming a problem is ill posed and a system that is likely not identifiable, and given this situation, we are trying to cope.  Therefore, we are not really solving an identifiability task because we are not trying to find \emph{the} best or most representative model that admits \emph{unique} parameter estimates; rather we are solving a problem more akin to, but not literally, a regularization task. We are starting from a point where the problem is both brutally ill-posed and likely non-identifiable, and where investigating identifiability using analytic methods, or even many numeric-by-hand methods are intractable. In this case we are assuming there will be a few different parameter combinations that represent reasonable parameter estimates. In this situation each combination of parameter represents a hypothesis for how the system works. More importantly, the method we present here is a flexible entry-point for using data assimilation with a complex nonlinear model and data collected in an uncontrolled environment rather than directly solving an identifiability problem.

\section{Data cohort}
\begin{table}[!ht]
% \begin{adjustwidth}{-1.25in}{0in} % Comment out/remove adjustwidth environment if table fits in text column.
\centering
%need to update demographics table for this particular data
\small
\begin{tabular}{|l|c|c|c|c|}
\hline
  \multicolumn{5}{|c|}{Data Summary} \\
\hline
Participant ID & P1 & P2 & P4 & P5 \\ \hline %\hline
Age & $40-50$ & $40-50$ & $40-50$ & $40-50$\\ \hline
Disease Status & T2D & T2D & No Diabetes & No Diabetes \\ \hline
Medications & metformin & metformin & --- & ---  \\ \hline
Total $\#$ glucose measurements &  $304$ &  $211$ & $520$ & $322$ \\ \hline
Total $\#$ meals recorded &  $124$ &  $76$ & $370$ & $184$ \\ \hline
Total $\#$ days measured & $16$ & $16$ & $53$ & $52$ \\ \hline
Mean measured glucose & $113 \pm 25$  & $127 \pm 32$  & $92 \pm 17$  & $101 \pm 16$ \\ \hline
% $\#$ of glucose measurements &  $304$ &  $211$ & $520$ & $325$ & $520$ & $325$ \\ \hline
% $\#$ of meals recorded &  $124$ &  $76$ & $370$ & $186$ & $370$ & $186$ \\ \hline
% %$\#$ of HbA1c measurements &  $2$ &  $1$ & $0$ \\ \hline \hline
% $\#$ of days measured & $27$ & $28$ & $92$ & $91$ & $92$ & $91$ \\ \hline
% Mean measured glucose & $112 \pm 25, (\sigma)$  & $129 \pm 33, (\sigma)$  & $91 \pm 16, (\sigma)$  & $102 \pm 17,  (\sigma)$ & $112 \pm 25, (\sigma)$  & $129 \pm 33, (\sigma)$ \\ \hline
\end{tabular}
\caption{Demographic information and summary statistics are reported for the four participants whose retrospectively collected data are included in the study. }
\label{table:participants}
\end{table}

We test and evaluate the Houlihan methodology in the context of modeling and forecasting blood glucose collected in a free-living setting --- via a type 2 diabetes self-management moblie application. The dlood glucose and nutrition data used here were collected retrospectively from four participants, two with type 2 diabetes and two without diabetes, using custom-designed mobile applications for capturing self-monitoring data (\cite{me_lena_1}). These data are summarized in Table~\ref{table:participants}.
We acquired two types of data: 1) fingerstick blood glucose measurements taken at the discretion of each of the 4 participants (roughly 3-10 times per day) and 2) estimates of carbohydrate consumption over time (roughly 1-5 meals per day) determined by a certified dietitian's analysis of the daily meal logs (with photos and descriptions) reported by each participant.The data are documented more completely in (\cite{da_glucose_forecast_t2d,online_offline_DA}) and are available on PhysioNet upon request.

\section{Methods}

\subsection{Glucose-insulin physiologic model}

The Houlihan method was conceived in the context of DA with a mechanistic model, and while it could be used in any nonlinear regression context, this paper will be restricted to the setting where we begin by projecting data onto a mechanistic dynamical system and then work to decide which parameters of that dynamical system we should estimate to represent the data. The mechanistic model is more formally either a dynamical system when time is discrete or a system of ODEs when time is continuous. Explicit versions of such systems form parameterized families of functions that are physically meaningful but generally do not satisfy nice function space properties such as completeness and are not universal approximators. The more general theory of dynamical systems can be found in many books (\cite{brin_ds_book,guck,arrowsmithandplace}), but here we will restrict our use of these details to an absolute minimum. We will assume that the systems we use have at least one invariant density; the invariant density is likely defined relative to a SRB-measure (\cite{youngSRB,ruelle_srb1,sinai_srb_1,bowen_srb1}) rather than Lebesque measure, but the point is that for a given set of parameter values and initial conditions, the states have a probability density function associated with them denoted $\Lambda$. \emph{This invariant density can potentially depend on both the parameters and the initial conditions for a set of parameters.} %For example, for the dynamical system $f(x_t)=ax_{t-1} mod_b$ with $a>1$ and $b>a$ will have different $\Lambda$'s for $x=0$, $x>0$, and $x<0$. Similarly, for the same system if $a<1$ then $0$ will be the only $\Lambda$ regardless of the initial conditions.

As previously noted, we want to use DA to model the glucose-insulin system of a human being.  We begin with a particular mechanistic glucose-insulin model, here the ultradian model that has been detailed in \cite{sturis_91,keenerII,pop_phys,dyn_pheno,da_glucose_forecast_t2d}, and has $6$ states and $21$ parameters; its details can be found in the appendix \ref{app:ultradian}.  The model has unknown identifiability properties, especially when only glucose is measured, but we have strong evidence that at least some of the model parameters and states are not identifiable (\cite{online_offline_DA}). The Houlihan method rests on quantifying how the invariant densities of the \emph{synthetic data sets} and their properties vary as parameters of the mechanistic model(s) vary. Specifically, the Houlihan method decides which parameters to estimate by varying the parameters of the ultradian model, observing how the invariant densities and their properties vary, and then using this information to select parameters to estimate by ranking ordering their importance using statistical inference or machine learning. The synthetic data used to select parameters to estimate will be generated by solving the ultradian model using an adaptive version of Runga-Kutta, ode23 in Matlab and will consist of $10^5$ simulated data points.

\subsection{Stochastic filtering and inverse problems methods}

We use two previously documented data assimilation formulations, an unscented Kalman filter (\cite{ukf_review,ukf_o,wan_primary_ukf,dual_ukf,wan_dual_ukf_wins,kalman_nn_ukf}) (UKF) whose details can be found in \cite{da_glucose_forecast_t2d} and a Metropolis-within-Gibbs Markov Chain Monte Carlo (MCMC) method whose details can be found in \cite{online_offline_DA,cotter_mcmc_faster}. As previously mentioned, these DA methods are used with the ultradian model (\cite{sturis_91}) for performing the DA tasks. We only use these methods over the course of evaluating the Houlihan methods; the exact implementation of the DA methods can be found in \cite{da_glucose_forecast_t2d,online_offline_DA}.

\subsection{Analytical construction and intuition for throwing the Houlihan around the right parameters}

\begin{figure}
\centering
\includegraphics[scale=0.16]{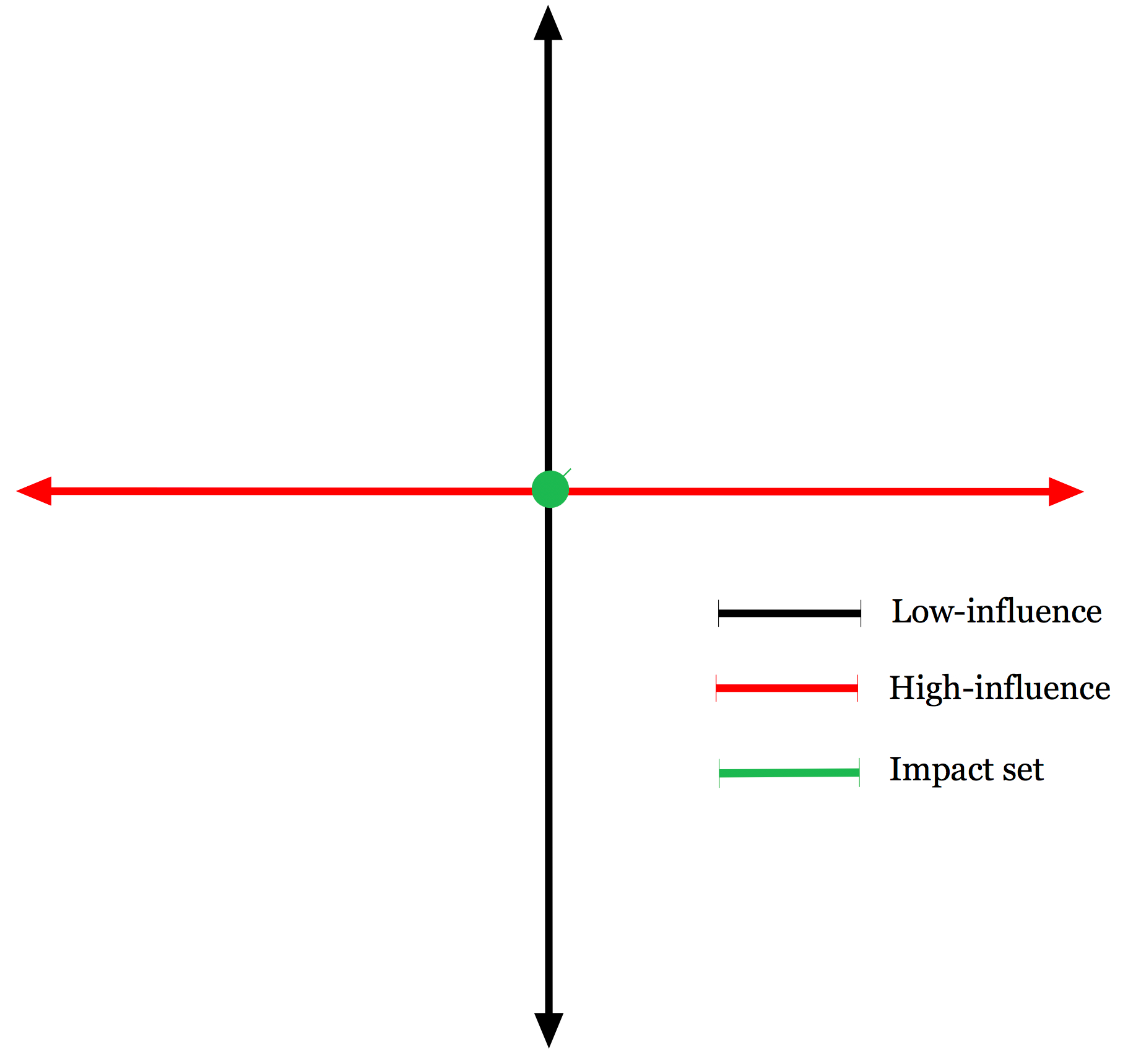}
\includegraphics[scale=0.16]{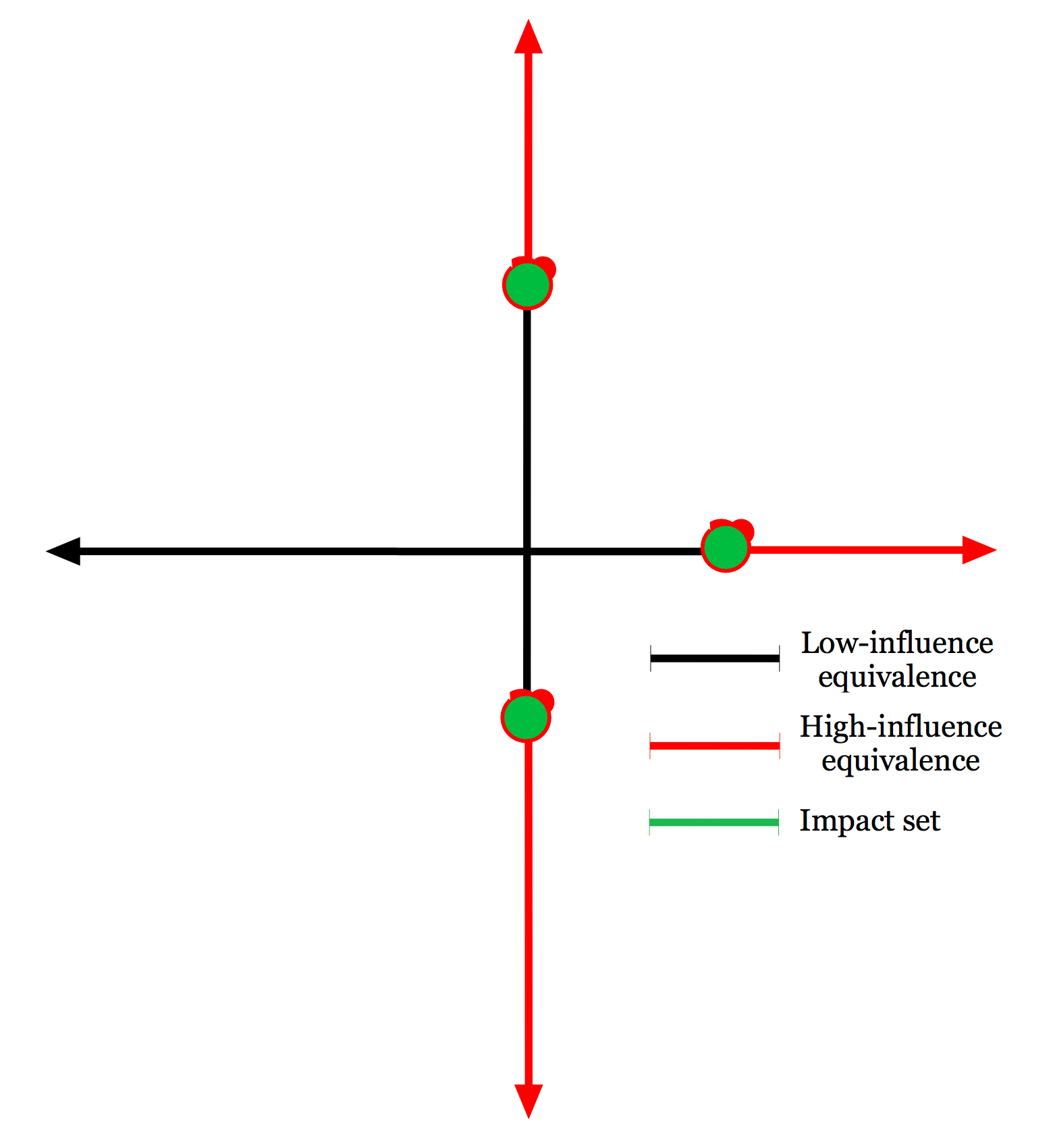}
\includegraphics[scale=0.16]{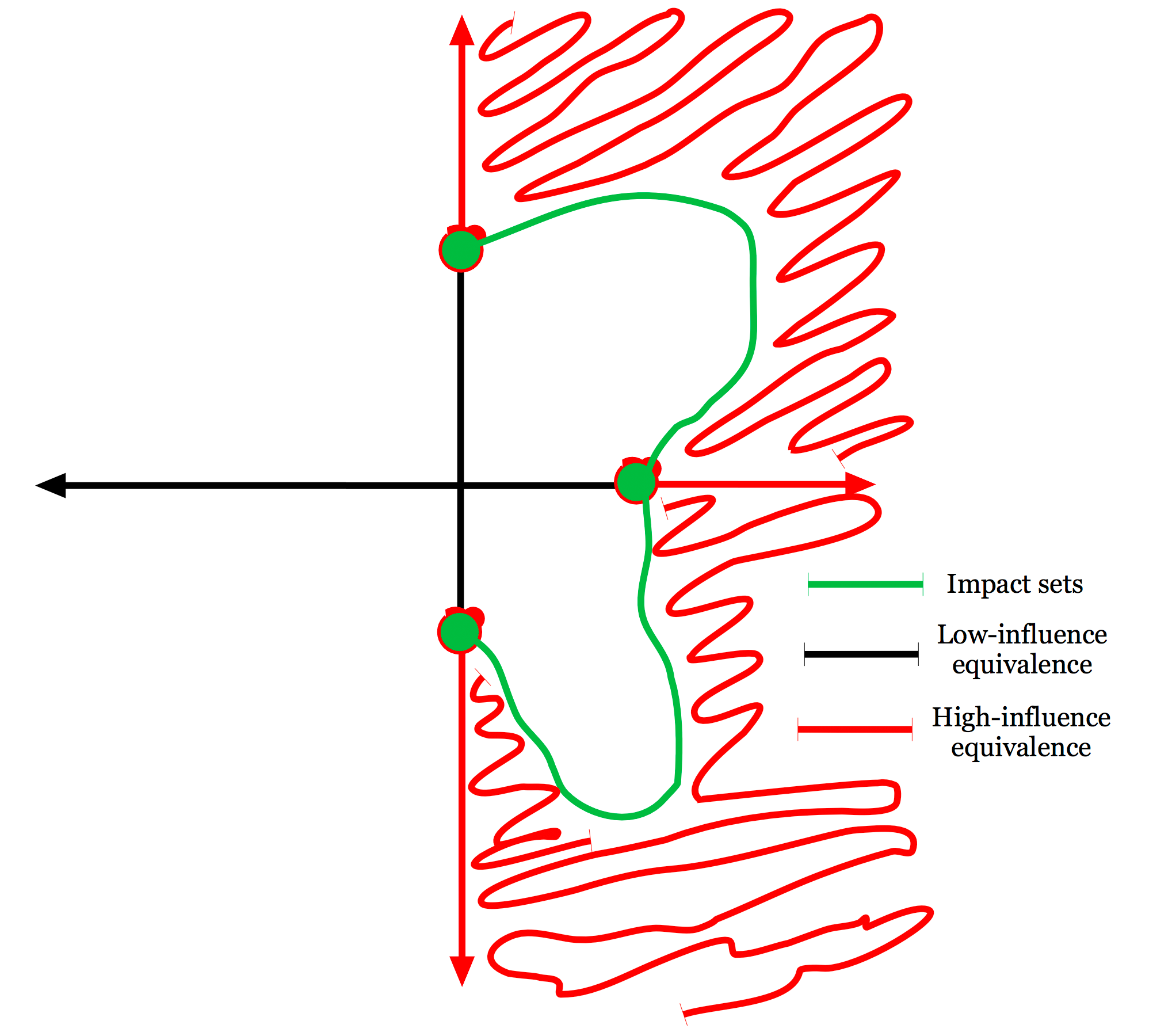}
\caption{Shown are three different Houlihan constructions: left shows equivalence class by coordinate---this is the construction we use in this paper; middle shows equivalence by subsets of coordinates but retains the non-joint parameter dependency assumption; right shows a fully joint equivalence where combinations of parameters can generate influence when individual parameters do not, similar to the notion of bifurcation sets.}
\label{fig:houlihan_fig}
\end{figure}

While the approach we are proposing is new, the \emph{allegorical} geometric intuition motivating this approach comes from bifurcation theory and in particular the bifurcation sets defined in the 1970's (\cite{soyomayorbifsets}) and the analytic geometry vision of bifurcation theory and singularities in parameter space \cite{arnoldgeo}. Bifurcation sets are the low-dimensional sets or manifolds that denote transition/bifurcation surfaces between topologically equivalent invariant sets, \emph{partitioning the parameter space into a set of equivalence classes.}  It is this idea of partitioning the parameter space into equivalence classes that differently impact dynamical featuers we care about is they key motivational insight. In our context we want to partition the parameter space by influence on some feature or set of features, denoted the \emph{feature-metric}, of the dynamics. Feature-metrics are calculated from the time-series of the simulated model (dynamical system), e.g., a mean. We do not want to be as rigid as requiring topological equivalence as was defined in the bifurcation sets framework, or necessarily strict classes, but we do want to partition the parameter space according to how parameters influence a dynamical feature we care about. The over-arching idea is that the subsets of parameter space that have the highest influence on the feature-metric are the parameters that will be the most useful to estimate to minimize Eq. \ref{eq:min}. And, knowing the most useful parameters to estimate provides a systematic way of choosing the parameters to estimate until the system is either identifiable or identifiable enough to be serviceable; in practice serviceable might mean that the errors are within desired tolerances, that parameter estimates are unique, or that the parameter estimates have few enough equilibria or minima that they can be made useful. To make this more precise, begin with the following terms, which are functions of a parameter vector, $p$.

\noindent
\emph{Feature metric}: the feature of the dynamical system we wish to influence, denoted $g(p)$; feature metrics are estimated from the time-series of the simulated model output and vary with parameter variation.
% \emph{Feature-metric}: the feature of the dynamical system we preserve, or the feature we use to define equivalence classes of parameters, denoted $\mathcal{F}$.

\noindent
\emph{Influence}: the amount that a parameter influences the feature-metric, denoted $F_i(g(p))$ for the i-th parameter.

% \noindent
% \emph{Influence function:} the property, or dependent characteristic of the feature-metric, $\mathcal{F}$, that is used to rank the influence of parameters.

\noindent
\emph{Influence equivalence}: a rule that defines equivalence of influence, e.g, all parameters $i$ such that $a_j \leq F_i(g(p)) < a_{j+1}$. This allows for us to introduce a partition over influence, $\{a\}_{j \in J}$ called an \emph{impact set}, which represents the transitions or boundaries between influence equivalence classes.

\noindent
\emph{Parameter influence sets}: the sets of parameters with equivalent influence according to the influence equivalence rule.

% \noindent
% \emph{Impact sets}: the transitions between sets in parameter space with different influence equivalence classes.

\textbf{Demonstrative example:} Begin by defining the dynamical system $f$ with state variables $x_i$ and parameters $p_i$ assuming at least four parameters. Next define the \emph{feature-metric} as the mean of a single state variable $x_{*}$, $\mu_{x_{*}}$ (i.e. we are interested in how each parameter "influences" the state's mean). Set the \emph{influence function} to be the absolute linear correlation, $|\beta_i|$, between the feature-metric, $\mu_x$, and values of the parameter $p_i$. In this example, the influence function is a vector-valued function, with a scalar metric (linear correlation between parameter and the state's mean it induces) corresponding to each parameter. The influence per parameter defines a probability mass function (PMF) with support $[0, 1]$ with values $\frac{|\beta_i|}{\sum_j{|\beta_j|}}$. Finally, we define \emph{influence equivalence} as membership in a given quartile of the PMF defined by the influence function.  Note that the \emph{impact set} is defined by the PMF quartiles, and the \emph{influence sets} are the parameters in respective quartiles of the PMF. Depending on the separation observed in the impact sets, we could ultimately choose to estimate parameters only from the upper equivalence class(es); i.e. the set of parameters with $|\beta_i|$ in the upper quartile. $\Box$

This example takes a narrow interpretation of the flexible construct we develop for identifying equivalence classes of parameter influence. However, even the above example allows for wild topological variation within a given equivalence class. For example, within a given equivalence class one would easily imagine there being many topologically distinct invariant sets due to \emph{both} parameter variation and initial condition variation. Presumably there are other similar equivalence class violations such as ergodicity properties (\cite{p-s-ergodic-attractors,burns_wilk_annals}), $k-LCE$ stability (\cite{hypviolation,dynamicsPRL}), etc. These issues can all be addressed by defining the various properties, e.g., the influence function, differently, or more restrictively such that we end up with increasingly more restrictive constructions such as the original notion of bifurcation sets. This flexibility in equivalencies is the point of this construction: we can, depending on our goals, data, etc., have substantial flexibility in how we set up how to choose what parameters to estimate all while explicitly acknowledging \emph{what we know we do not know we are preserving.} For example, if we define the feature-metric to be the mean, we know we are allowing the system to explore or have many different coexisting invariant densities as long as they have a mean that lies within a given equivalence class.

\textbf{Visual example:} Figure \ref{fig:houlihan_fig} shows three cases of the outcome of the Houlihan analysis.  The left-most plot in fig. \ref{fig:houlihan_fig}  shows the case where the rank-ordering of influence is on a by-coordinate basis; meaning, the equivalence classes were collections entire coordinates, here where each equivalence class has a single member.  The middle plot in fig. \ref{fig:houlihan_fig} shows a case where the influence can be portions of different coordinates, but still there are is not joint dependence between variables.  The right-most plot in fig. \ref{fig:houlihan_fig} demonstrates an example where the influence equivalence includes joint coordinate relationships. In this paper we will only address the first of these cases, leaving the more complex situations for later work.

\subsection{Computational moving parts for throwing the Houlihan around the right parameters}

The computational task of selecting parameters to estimate involves defining the equivalence-like classes, finding their boundaries, rank ordering the parameters by importance and has, broadly, five moving parts.  \emph{First}, select the feature-metric(s), $g(p)$, e.g., mean. \emph{Second}, formulate the representation of the space of parameters and their variation, including (i) parameter grid resolution, (ii) parameter perturbation range, (iii) parameter variation type, e.g., joint versus individual by-parameter parameter variation. \emph{Third}, choose an influence function that defines how to model the parameterized variation of the feature-metric variation with parameter variation. \emph{Fourth}, choose a method for rank ordering these parameterizations by influence. Sometimes steps three and four can be done using a single method, e.g., linear regression with a $L_1$ regularization or by using lasso with cross validation, and sometimes it is done in two steps, e.g., linear regression with a threshold on the $\beta$'s, partitioning the $\beta$'s into equivalence classes. And \emph{fifth}, decide which parameters to keep or which equivalence classes, or which impact sets are important.

%We will evaluate the method by evaluating how perturbations of these choices changes the forecast errors and the ability to resolve parameter estimates.

\paragraph{Feature metrics} We use two feature metrics, mean and standard deviation of the invariant density generated by mechanistic model with set parameter values and initial conditions.

\paragraph{Parameter grid} We begin with the nominal parameters (\cite{sturis_91,keenerII,da_glucose_forecast_t2d}), and then vary them in intervals of $\log_2$ over $10$ decades in both directions. For example, for parameter $i$ the parameter grid point for the $k^{th}$ decade was set as $p_i(nominal) 2^k$. We did not consider joint-variation of parameters, but varied parameters independently while holding all other parameters fixed at their nominal values.

\subsubsection{Parameter selection methods: Influence functions, impact sets, and ranking}

Given a feature metric as a covariate or input vector, e.g., the means of attractor densities for a set of parameter values,  we use several methods for selecting the best set of parameters to estimate in a DA context.  Some of these methods are stock---linear regression with lasso---some are standard practice---parameter selection using knowledge of the model---and some are modifications of existing methods---see PCA-lariat below. We will see that the method for selecting the parameters matters, although not as much as the feature metric, and it is clear that sophisticated machine learning methods could be useful in this context.

\paragraph{Covariates or input vectors} All of the methods below take a covariate matrix as input.  The covariates correspond to vectors: one dimension of the covariate matrix corresponds to a feature metric calculated at every point along the parameter variation, e.g., the mean of a simulated attractor at every point along a one-dimensional parameter curve.

 \paragraph{By hand selection parameter selection --- parameter selection using knowledge} In our previous work we selected parameters to estimate by hand as they were tied to certain dynamical features, physiologic knowledge want to fit something in particular to solve a problem, e.g., phenotyping. We selected $E$ and $V_p$ because they seemed to have an impact on the mean (\cite{dyn_pheno}) and $t_p$ because it was related to liver function; the results can be found in \cite{da_glucose_forecast_t2d}.

\paragraph{Automatic parameter selection using linear regression}

A basic method for determining influence is the linear dependence between the feature metric and parameter variation. In this setting we perform a linear regression between the feature metric and the parameters and we keep all $\beta$'s for which $\beta_i  > (\beta_{1}) (\kappa_{LR})$.  Here we set $\kappa_{LR} = 20 \%$ or $0.2$, meaning that we keep all the parameters that have a regression coefficient that explains at least $20 \%$ of the regression coefficient with the highest influence.

\paragraph{Automatic parameter selection using Lasso and cross validation}

A natural way of reducing the number of parameters in a model is to select parameters that have a lot of power explaining the feature metric while simultaneously being non-redundant.  One way of achieving this is to use lasso, or $L_1$ regularization to enforce a sparse representation of the parameter system (\cite{statistical_learning_sparsity,lasso_1,lasso_2,elastic_nets}). We use the standard lasso formulation (\cite{statistical_learning_sparsity}) with cross validation to determine the rank-ordering of parameters; the optimal value of $\lambda$, or the optimal number of parameters, is set using a cutoff of one standard error.  Lasso automatically and uniquely rank orders parameters.  We keep the parameters within one standard error of the minimum mean squared error (MSE) ensuring a sparse representation of the model.

\paragraph{Automatic parameter selection using elastic net approximation of ridge regression}

In addition to lasso regularization, we also use ridge regression, or $L_2$ regularization (\cite{statistical_learning_sparsity,elastic_nets,elastic_nets_are_linear_svm_1}). We compute the ridge regression selected parameters using an elastic nets formulation with $\alpha$ set to $0.0001$ where elastic nets formulation approachs $L_2$ regularization, and select the number of parameters using cross validation in the same way as is done in the lasso setting. We keep the parameters within one standard error of the minimum mean squared error (MSE) ensuring a sparse representation of the model.

\paragraph{Automatic parameter selection using PCA-lariat with a single metric} To add diversity to the set of methods for selecting parameters beyond linear regression-based methods, we devised a principle component analysis (PCA) (\cite{pca_original,pca_original_2,pca_book}) based algorithm for computing an influence function, then implement a rank-ordering scheme for defining influence equivalence. The method we develop, \emph{PCA-lariat}, follows seven steps. \emph{First,} estimate the PCs for the feature-metric, $g(p)$, taking care to de-trend the summary. \emph{Second,} estimate the percentage of the variance captured by the $i-th$ PC, $\sigma_{PC}(i)$. \emph{Third,} identify the \emph{important PCs}, or the PCs that explain variance above a threshold, $\kappa_{PC}$; we use $5 \%$. \emph{Fourth,} for each important PC, rank-order the contribution of each parameter or coordinate to the PC. \emph{Fifth,} collect all the coordinates for all the important PCs that contribute proportionally to a given PC above a set threshold, $\kappa_{C}$, $PC_j(i)>\kappa_{C}$; we use $10 \%$. \emph{Sixth,} for the important parameters for the important PCs, estimate the contribution per parameter:
\begin{equation}
PCR(i) = \sum_j \sigma_{PC}(j) * PC_j(i).
\end{equation}
And \emph{seventh,} rank order the important parameters by $PCR$ and select the parameters above a given threshold, $\kappa_{I}$; we use $0.1$, or $10 \%$.

%Maximal: union of the top X parameters across evaluation metric types (e.g., mean, variance).
%Minimal: intersection of the top X parameters across evaluation metric types (e.g., mean, variance).

\paragraph{Multi-directional parameter wrangling} Combining models, or model averaging can be very useful for improving results (\cite{bayesian_model_average_mad,model_average_calibrate_aos,claeskens_model_selection,da_glucose_forecast_t2d}), especially when you either know you want to adjust to multiple feature-metrics, or you do not know what feature metrics are important. Here, we only consider using set operations over methods, and consider three cases. First, we take the union of: (number of rank-ordered parameters, feature-metric, influence function) using one parameter per influence function, two feature metrics, mean and standard deviation.  Second and third, we take the union of: (number of rank ordered parameters, feature-metric, influence function) using one parameter per influence function and one feature metric, either mean or standard deviation.

%Maximal: union of the top X parameters across evaluation metric types (e.g., mean, variance).
%Minimal: intersection of the top X parameters across evaluation metric types (e.g., mean, variance).

\subsection{Evaluation scheme}

The evaluation of the Houlihan methods is done in four steps. \emph{First}, we apply the Houlihan methods to the ultradian model to select parameters to estimate and compare the parameter selections as the method is perturbed.  \emph{Second}, we use both the UKF and the MCMC DA methods to estimate these Houlihan-selected parameters for the four people in our cohort and calculate the mean squared error (MSE) between the data and the model state estimates (MCMC methods) and forecasts (UKF methods). \emph{Third}, we use both the UKF and the MCMC DA methods to estimate parameters for \emph{both} parameters that were previously chosen by hand in previously published work (\cite{da_glucose_forecast_t2d}) and parameters that the Houlihan methods determined were low-influence parameters and again calculate the MSE between the data and model state estimates and forecasts. \emph{Fourth}, we compare the MSE for the variously selected parameter sets.

\section{Results}

The results come in two stages.  \emph{First}, we present the rank-ordered parameters selected by different methods in order to demonstrate: (i) which parameters the methods selected, (ii) that the methods selected some but not all parameters, (iii) how the parameter selection varied across methods, and (iv) the rank-ordering of parameters by method. \emph{Second,} we evaluate the methods by using the parameters selected in each method to forecast glucose with the UKF and smooth glucose with MCMC; methods are compared via the MSE between measurements and predictions.

\subsection{Parameter selections by method}

% \begin{tabular}{ | l | l | l | p{5cm} |}
\begin{table}
\centering
\small
%\resizebox{\columnwidth}{!}{
\begin{tabular}{ |l|l|l|l|l|l|l|l|l|l|l|l|}
  \hline
  \multicolumn{12}{|c|}{Rank-ordered parameters per selection method out of $21$ possible parameters} \\
  \hline \hline
  method & $1$ & $2$ & $3$ & $4$ & $5$ & $6$ & $7$ & $8$ & $9$ & $10$ & $11$ \\ \hline
  LASSO  $\mu$ &  $a_1$ & $C_1$  & $V_p$  & $t_p$   & $R_m$ & $C_3$ & ---  & --- & ---& ---& ---\\ \hline
  LASSO  $\sigma$ & $R_g$ & $C_3$  & $U_m$  & $a_1$   & $C_1$ &  $t_p$ & $R_m$ & $V_p$ &  -- & --- & ---   \\ \hline
  Linear regression $\mu$ &  $a_1$ & $C_1$  & $C_3$  & $R_m$ & $t_p$ & $V_p$ & $U_m$  & $R_g$ & $C_4$ & $U_b$ & $U_0$  \\ \hline
  Linear regression $\sigma$ & $R_g$ & $C_3$  & $U_m$ & $a_1$  & $C_1$ & $R_m$ & $V_p$ & $t_p$ & $k_{decay}$ & --- & ---    \\ \hline
  Ridge regression $\mu$ & $a_1$ & --- & --- & --- & --- & --- & --- & --- & --- & --- & ---  \\ \hline
  Ridget regression $\sigma$ &  $R_g$ &  --- & --- & --- & --- & --- & ---  & --- & --- & --- & --- \\ \hline
  PCA $\mu$ & $a_1$ & $C_1$ & $C_3$ & $R_m$ & $t_p$  & $V_p$ & $U_m$ & --- & --- & --- & --- \\ \hline
  PCA $\sigma$ & $R_g$  & $C_3$ & $U_m$ & $a_1$ & $C_1$  & --- &  --- & --- & --- & --- & ---\\ \hline
   \hline
\end{tabular}
%}
 \caption{The rank ordering choice of the four parameter selection methods for the feature-metrics mean, $\mu$, and standard deviation, $\sigma$.}
   \label{table:whochosewhat}
\end{table}

% Fig. \ref{fig:rank_order} shows how the $l_1$, $l_2$ and PCA-based methods rank-order the parameters according to how they influence the mean.  Lasso has the advantage that it always adds parameters one at a time, but here all methods added parameters one at a time.
Table \ref{table:whochosewhat} shows the rank-ordered parameters selected by each parameter selection method. The methods were sensitive to the feature metric; the mean and standard deviation-based methods did not select the same parameters as important.

For a given feature metric, all selection methods identified the same top two parameters --- all methods ranked $a_1$ and $C_1$ as the top influencers of the mean, and ranked $R_g$ and $C_3$ as the top influencers of the standard deviation. However, the entire influence sets differed substantially. This indicates that influence set structure, as defined (upper quartile of influence), is sensitive to choices of influence functions and influence equivalence definitions.

\begin{figure}
%\begin{adjustwidth}{-2.25in}{0in} % Comment out/remove adjustwidth environment if table fits in text column.
\centering
\includegraphics[scale=0.2]{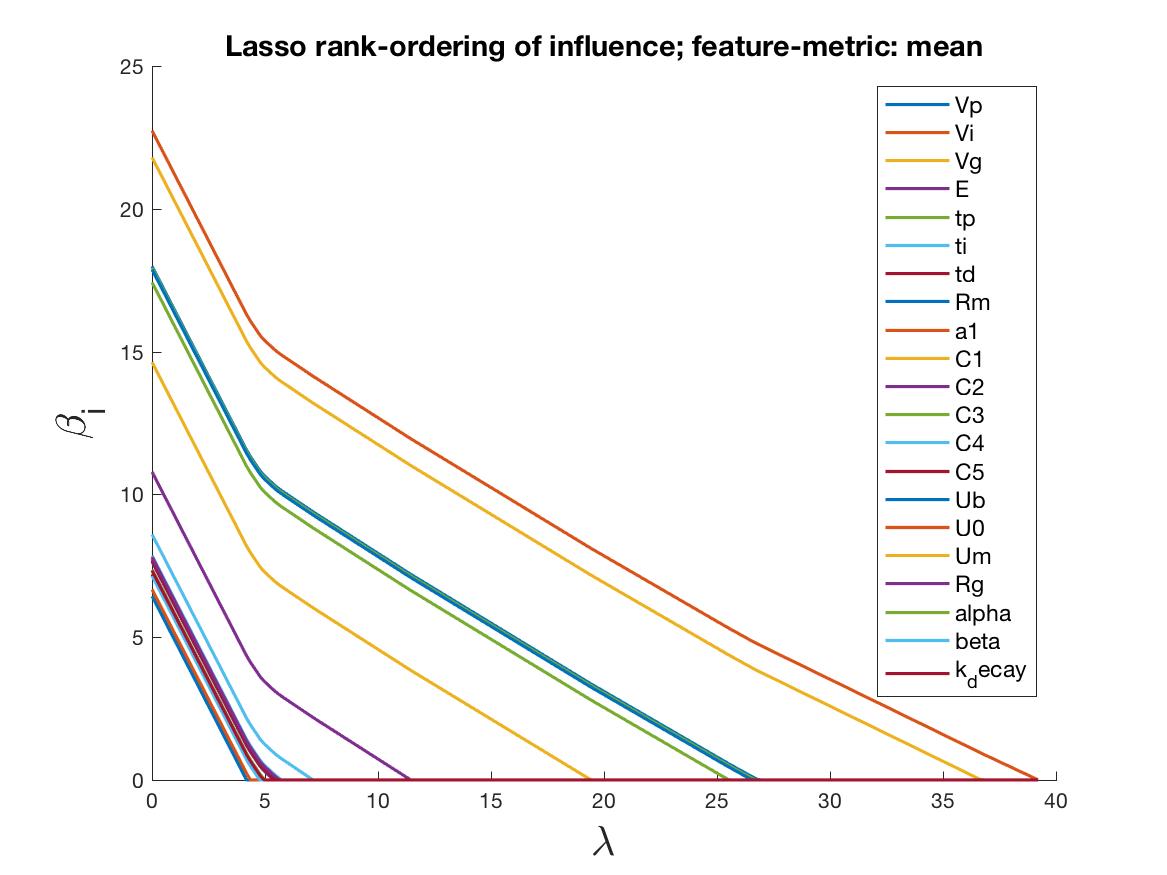}
\includegraphics[scale=0.2]{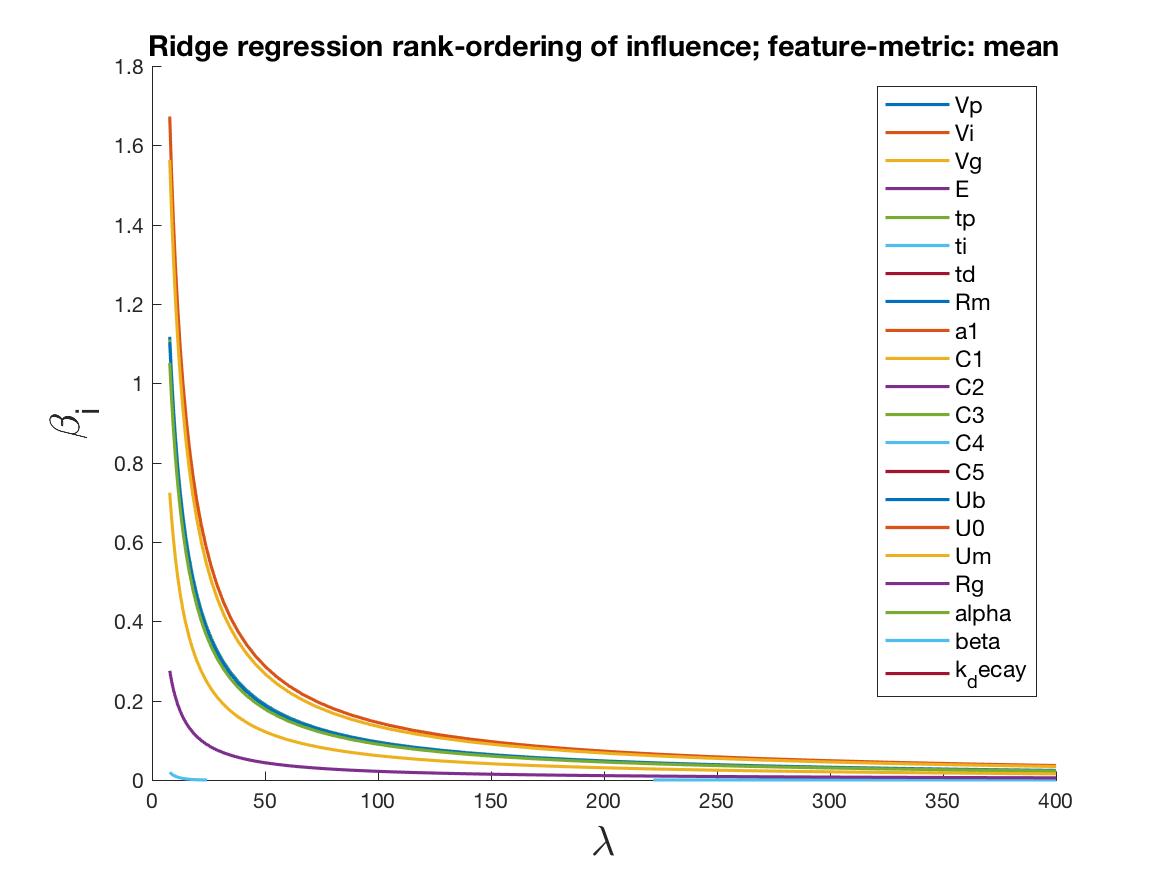}
\includegraphics[scale=0.2]{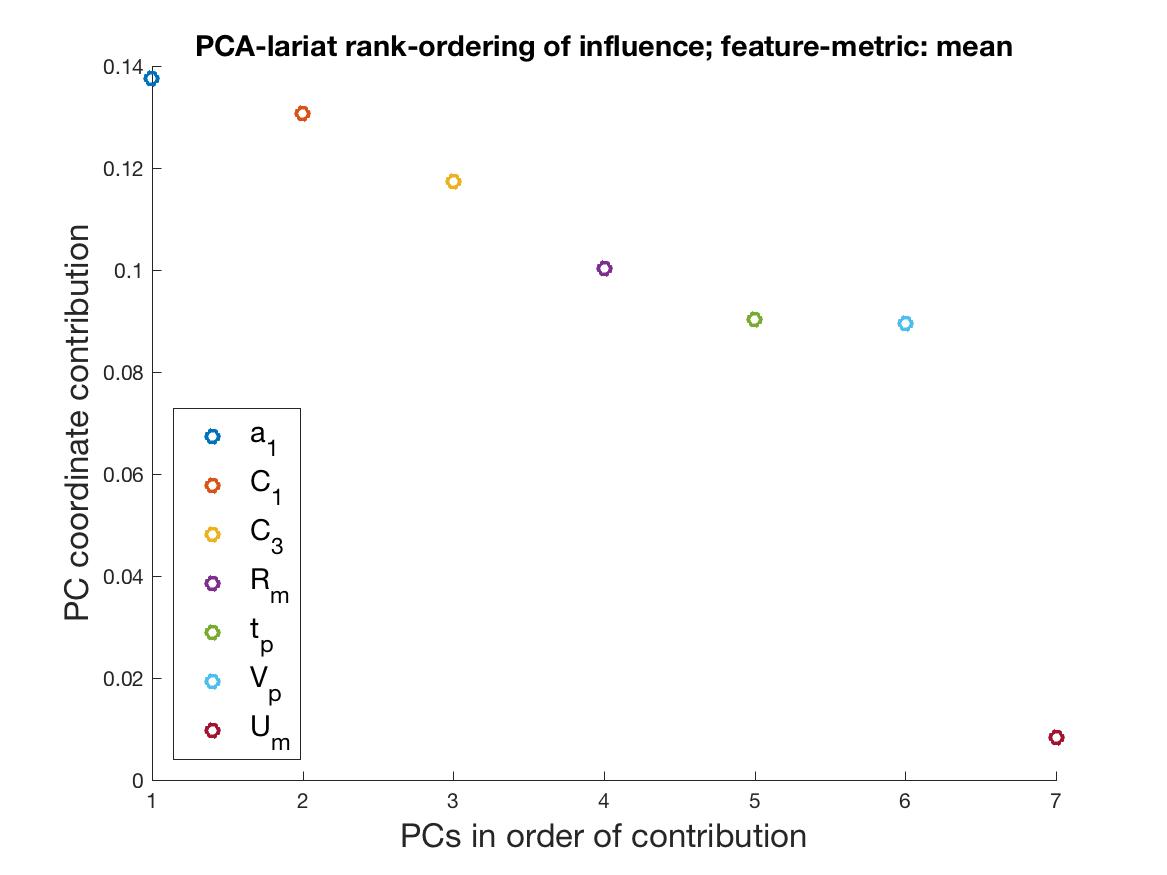}
\caption{The rank-ordered influence function with a feature-metric set to the mean for lasso, ridge regression, and PCA-lariat methods.}
\label{fig:rank_order}
%\end{adjustwidth}
\end{figure}

Interestingly, the equivalence classes of high and low parameter influence are preserved under perturbations to the influence function.
Fig. \ref{fig:rank_order} shows how the $l_1$, $l_2$ and PCA-based methods rank-order parameters according to how they influence the mean. While lasso is expected to preserve the ordering with different $\lambda$ (it fits one-at-a-time), ridge regression also remains robust to variations in the regularization term, $\lambda$, adding parameters one at a time.

Most methods find only $5-6$ influential parameters out of $21$, greatly reducing the dimension of the parameter space. In all cases, the methods gave an entry point for which parameters to begin estimating; the next question, then, is whether using the Houlihan approach helps to reduce forecasting errors and improve convergence of parameter estimates.

%From the results in  can be summarized as follows:
%\begin{enumerate}
%\item the rank ordering of parameters to estimate is sensitive to the metric-feature preserved, $\Phi$;
%\item the rank ordering of parameters to estimate is sensitive to the method used to estimate the importance of parameters;
%\item different regularization methods, e.g., LASSO versus ridge, for learn regression rank order parameters differently in non-trivial ways;
%\item different methods for estimating the importance of parameters identify substantial redundancy in the parameters --- most methods only find 4 or 5 useful parameters:
%\begin{enumerate}
%\item one parameter estimates: the eight methods select one of three parameters as the most important;
%\item two parameter estimates: the eight methods select one of three parameter 2-tuples as the most important;
%\item three parameter estimates: the eight methods select one of five parameter 3-tuples as the most important;
%\item four parameter estimates: the eight methods select one of five parameter 4-tuples as the most important;
%\end{enumerate}
%\end{enumerate}
%That the most important set of parameters to estimate depends on both the metric-feature being optimized against and the method for selecting the most important parameters is not surprising.  The primary question is, when we use these parameter selection methods work for choosing the parameters to estimate states, which method-metric-feature-pair achieves the best results.

%/Users/albers/Dropbox/high_throughput_identifability/src

\begin{figure}
%\begin{adjustwidth}{-2.25in}{0in} % Comment out/remove adjustwidth environment if table fits in text column.
\centering
\includegraphics[scale=0.18]{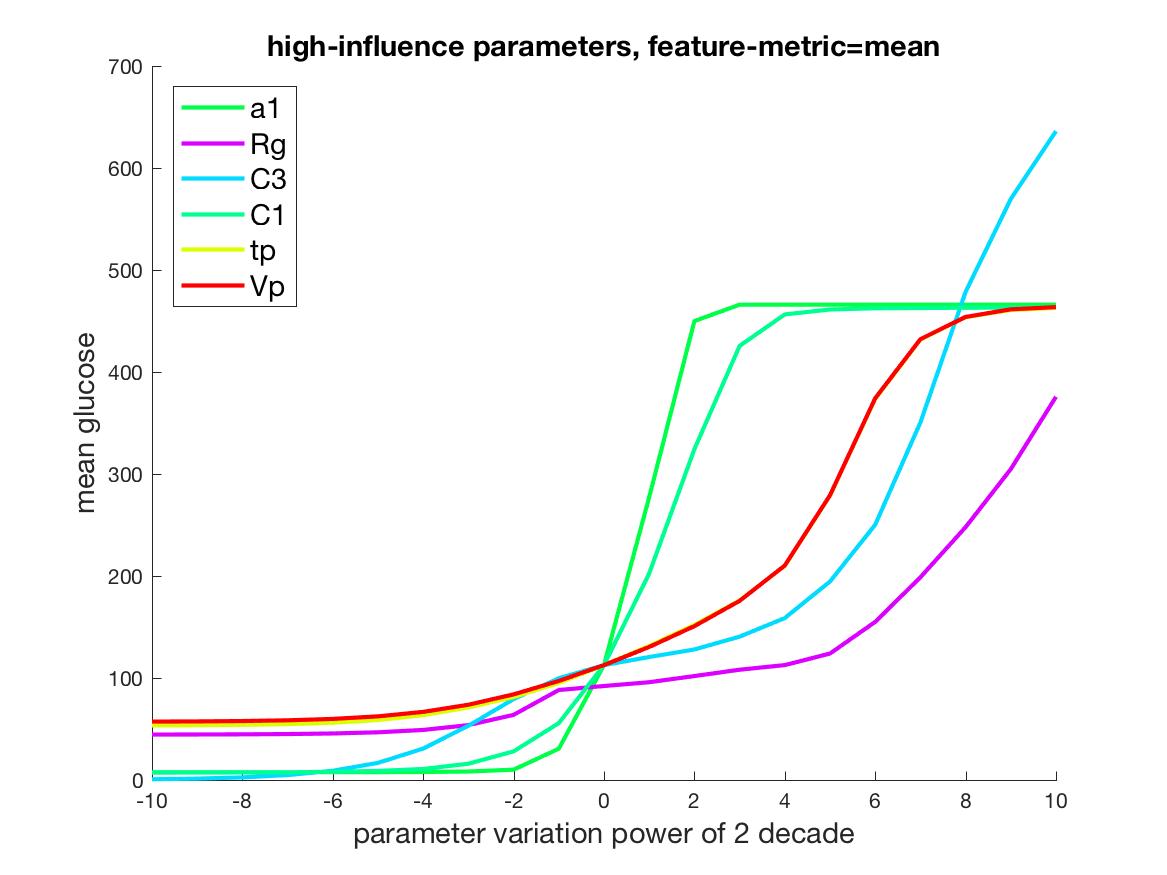}
\includegraphics[scale=0.18]{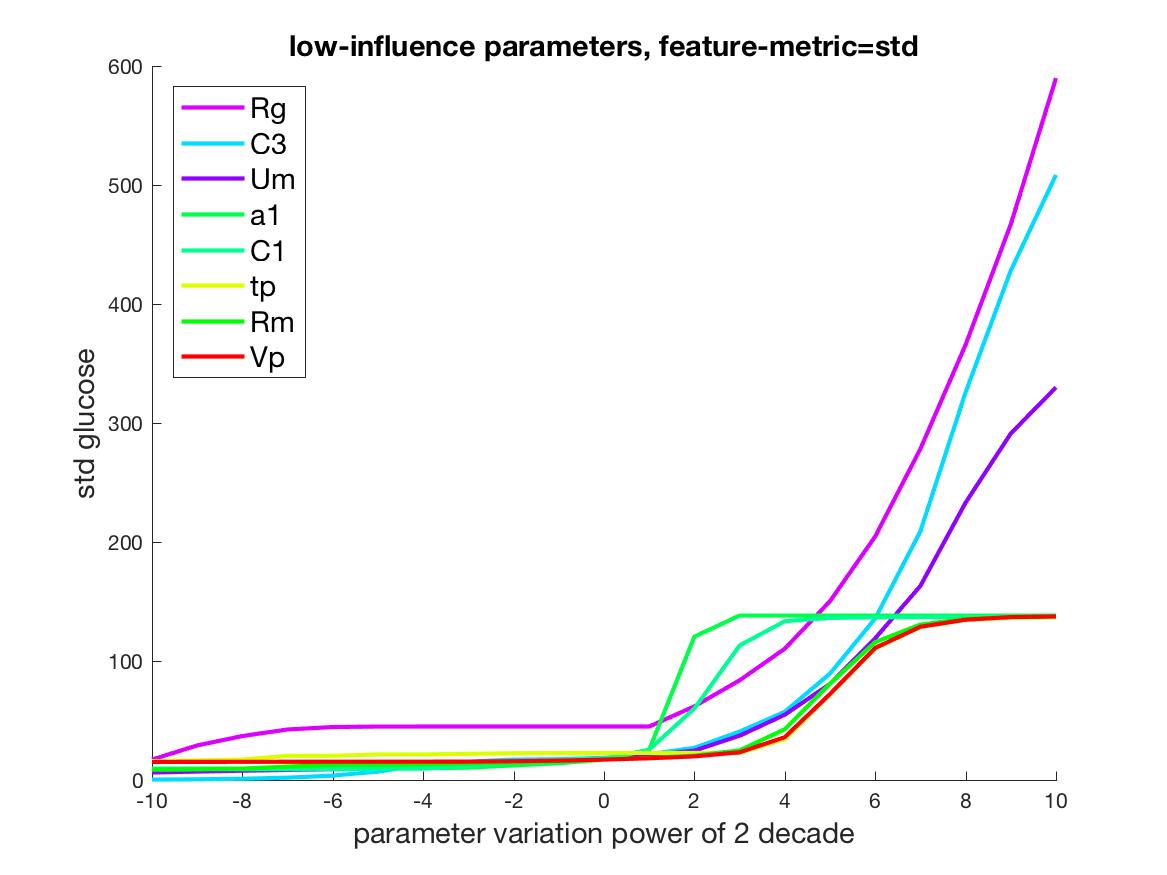}
\includegraphics[scale=0.18]{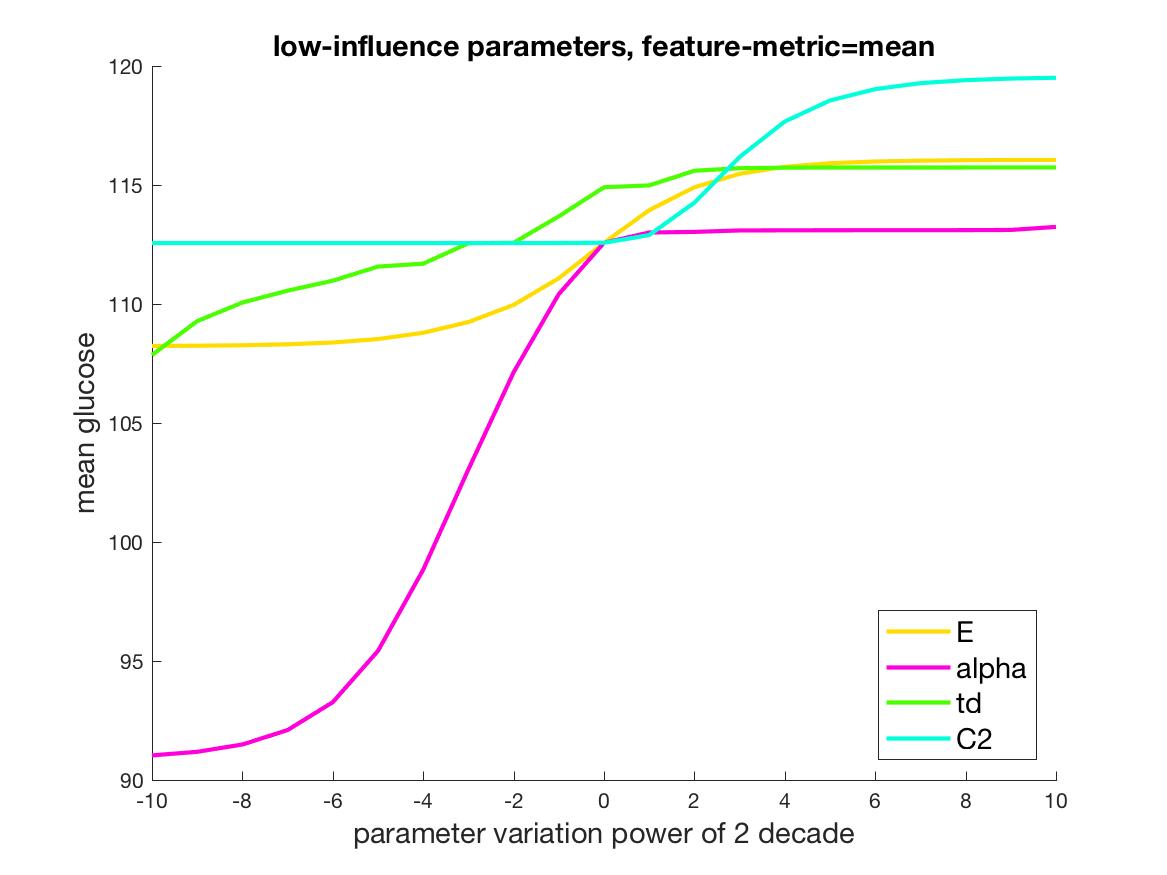}
\includegraphics[scale=0.18]{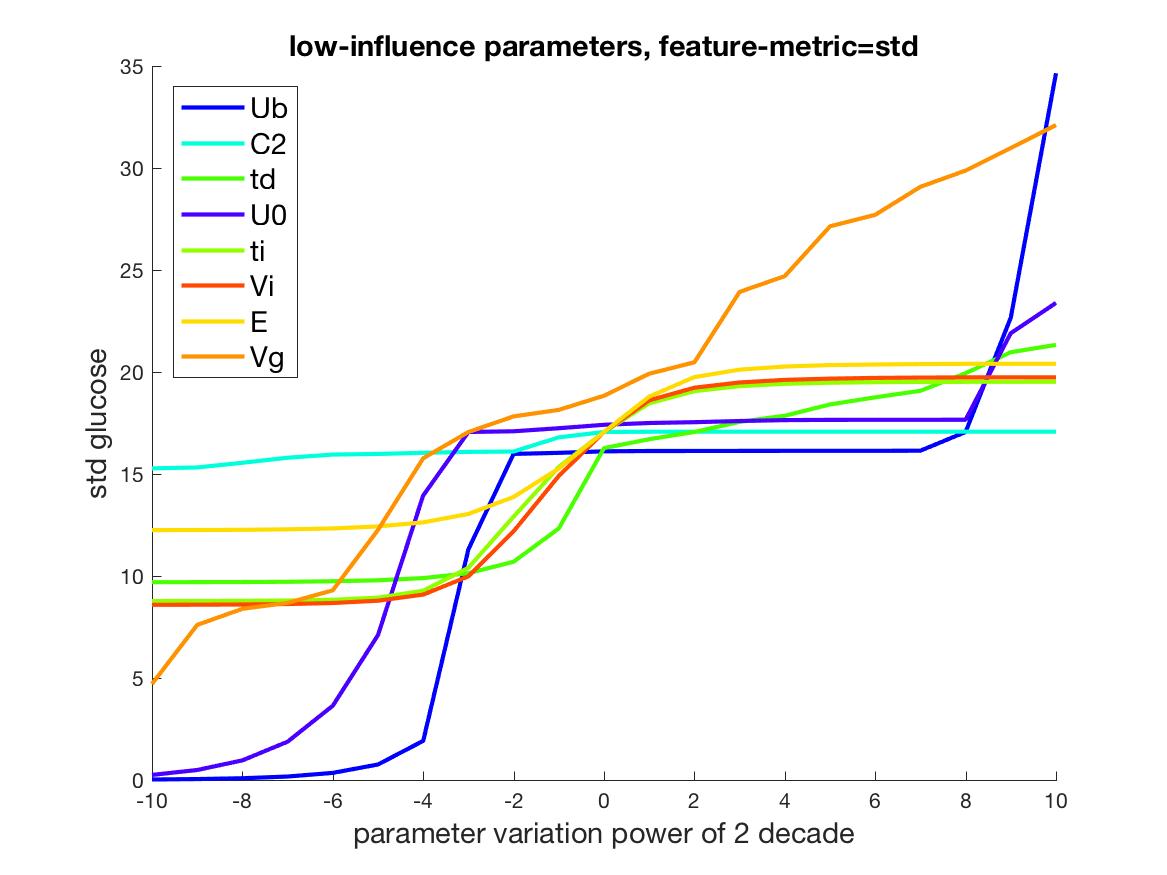}
\caption{The influence for two feature metrics, mean and standard deviation, versus parameter variation for high impact and low impact parameters.}
\label{fig:impact}
%\end{adjustwidth}
\end{figure}

\paragraph{Redundancy and influence} Our goal is to select parameters to estimate during forecasting and smoothing tasks. We aim to facilitate this goal by identifying small parameter sets that have significant, minimally redundant influence over important dynamical features. Accomplishing this can minimize problems in identifiability, multiple coexisting invariant sets, etc.
% Fig. \ref{fig:impact} shows how this works relative to the influence curves.
Fig. \ref{fig:impact} visualizes variation of the feature-metric, mean and standard deviation of the invariant density with parameter variation, as well as how the methods partitioned parameters into a high and low-influence equivalence class. It is clear that some variations in some parameters create large shifts in the mean and variance (e.g. $a_1$), whereas the mean and variance features are far less sensitive to other parameters, like $E$ and $t_d$.

While the mean and standard deviation are not always influenced by the same parameters, the methods select parameters that have both high influence and relatively orthogonal influence; e.g., in the case of the mean the methods generally select $a_1$ and $R_g$ first. The low influence parameters, by comparison, are not able to move the mean or standard deviation appreciably and are therefore not able to fully explore the space.  Similarly, the low influence parameters are relatively redundant. Following this logic one might predict that estimating alpha and $C_2$ would lead to the most accurate model estimates while estimating $E$ and $t_d$ would lead to the least accurate model estimates.

\paragraph{Comparison with by-hand selection} In our previous work (\cite{da_glucose_forecast_t2d}) we selected parameters to estimate by hand based on our desire to estimate certain parameters related to physiologic function, e.g., $t_p$, and because of their obvious influence on parameters, e.g., $V_p$ as could be deduced from other previous work (\cite{dyn_pheno}) to influence the mean state. The automated methods selected $V_p$ and $t_p$ as high influence parameters, but not $E$, a parameter the methods determined was a low-influence parameter.

%We will compare our previous by-hand-selected parameters to the Houlihan selected parameters. The rank ordering and correlation coefficients ($\beta$'s) for the mean and standard deviation respectively for the hand-selected parameters are as follows: $E$ has a ranking of $(11, 14)$, and regression coefficients of $(1.12, 0.16)$, has a ranking of   $V_p$ $(3,5)$ and regression coefficients of $(11.47, 3.62)$  and $t_p$ has a ranking of $(17, 8)$ and regression coefficients of LR$(0.73, 0.88)$; any ranking below fourth or fifth generally has very little impact or differentiation from other low-ranking parameters. Interestingly, $V_p$ and $t_p$ have nearly identical influence on the feature metrics, making them highly redundant parameters to estimate and likely exacerbating identifiability problems.

\subsection{Parameter selection method evaluation}

\begin{table}
\centering
\resizebox{\columnwidth}{!}{
\begin{tabular}{ |l|l|l|l|l|l|l|l|l|p{4cm}|}
  \hline
  \multicolumn{10}{|c|}{Rank-ordered parameters per selection method} \\
  \hline \hline
    & \multicolumn{4}{|c|}{MSE for MCMC} & \multicolumn{4}{|c|}{MSE for UKF} &  \\ \hline \hline
  parameter  & P1 & P2 & P4 &  P5 &  P1 & P2 & P4 & P5 & method-feature-metric pairs    \\ \hline
   $a_1$ & $822$ & $1140$ & $338$ & $296$  & $809$ & $1270$ & $304$ & $356$   & LASSO($\mu)$, LR($\mu$), PCA($\mu$) \\ \hline
   $R_g$ & $655$  & $1180$ &  $475$  & $288$  & $672$ & $1490$ & $470$ & $401$  & LASSO($\sigma$), LR($\sigma$) , ridge($\sigma$), PCA($\sigma$) \\ \hline \hline
   $t_p$  & $807$ & $1020$ & $448$ & $349$  & $788$ & $1050$ &  $407$ &  $420$ & by-hand, high-influence\\ \hline
   $V_p$  & $820$ & $1120$ & $332$ & $320$ & $805$ & $1300$ &  $313$ & $362$ & by-hand, high-influence\\ \hline
   $E$  & $681$  & $1250$ &  $655$ & $500$  & $721$  &$1380$ &  $704$ & $724$ & by-hand, low-influence\\ \hline \hline
   $\alpha$  & $501$ & $1250$ & $526$ & $346$  & $526$ & $1580$ & $528$ & $394$ & low-influence \\ \hline
   $t_d$  & $530$  & $1080$ & $730$  & $674$ & NaN & $$1260 &   NaN & $480$ & low-influence \\ \hline \hline
     \multicolumn{10}{|c|}{Rank-ordered parameter pairs per selection method} \\
  \hline \hline
     $(a_1, C_1) $  & $570$ & $1080$ &  $285$ & $276$  & $698$  & $1290$ & $258$ & $330$ & LASSO($\mu)$, LR($\mu$), PCA($\mu$)  \\ \hline
     $(R_g, C_3) $  & $593$ & $923$ & $210$ & $297$  & $613$ & $1260$ & $215$ & $385$ & LASSO($\sigma$), LR($\sigma$) , ridge($\sigma$), PCA($\sigma$)\\ \hline \hline
     $(a_1, R_g) $  & $578$  & $1130$ & $292$ & $296$   & $614$ & $1400$  & $269$ & $343$ & Union of rank $1$ over methods \\ \hline \hline
     $(\alpha, E)$  & $454$ & $1174$ &  $518$ & $345$  & $483$ & $1310$ & $535$ & $520$ & low-influence \\ \hline
     $(\alpha, t_d)$  & $432$ & $993$ & $525$  &  $347$  & NaN & $1120$ & NaN & NaN & low-influence \\ \hline
     $(E, t_d)$  & $462$ & $1030$ & $592$ & $487$  & $643$ & $1190$ & NaN & $490$ & low-influence \\ \hline \hline
     \multicolumn{10}{|c|}{Rank-ordered parameter $3$-tuple per selection method} \\ \hline \hline
     $(a_1, C_1, V_p) $  & $569$ & $1060$ & $284$ & $276$   & $663$ & $1310$ & $260$ & $329$ & LASSO($\mu$) ($1^{st}$) \\ \hline
      $(a_1, C_1, t_p) $  & $518$ & $864$ &$247$  & $275$  & NaN & NaN & $234$ &  $294$ & LASSO($\mu$) ($2^{nd}$) \\ \hline
     $(R_g, C_3, U_m) $  & $590$ & $922$ & $190$ & $294$ &  $618$ & $1140$ & $228$  & $391$ & LASSO($\sigma$),  LR($\sigma$),  PCA($\sigma$) \\ \hline
     $(a_1, C_1, C_3) $  &  $431$ & $1020$ & $261$ & $274$  & $1330$ & $1110$ & $251$ & $340$ &  LR($\mu$) PCA($\mu$)\\ \hline
     $(C_2, E, \alpha)$  & $442$  & $1020$ & $518$  & $346$ &  $479$ & $1250$ & $535$ & $515$ & low-influence \\ \hline
     $(t_d, C_2, \alpha)$  & $432$ & $894$ & $525$ & $347$  & NaN   & $1190$ & NaN & NaN & low-influence \\ \hline
     $(t_d, E, \alpha)$  & $398$ & $956$ & $479$ & $343$  & NaN & $1120$ & NaN & $520$ & low-influence \\ \hline
     $(t_d, E, C_2)$  & $464$  & $941$ & $592$ & $489$ & $630$ & $1190$ & NaN & NaN & low-influence \\ \hline \hline
       \multicolumn{8}{|c|}{Rank-ordered parameter $4$-tuple per selection method} \\ \hline \hline
       $(a_1, R_g, C_1, C_3) $  &  $398$ & $864$ & $182$  &  $288$ & $649$  & $985$ & $265$ & $324$ & Union of rank $2$ over methods \\ \hline
       \multicolumn{10}{|c|}{Full Houlihan for $\mu$ and $\sigma$ } \\ \hline \hline
       $(a_1, C_1, V_p,  t_p, R_m, C_3) $  &  $414$ & $862$ & $217$  & $229$  & $661$  & NaN &  $236$ &  $291$ & Lasso($\mu$)  \\ \hline
       $(R_g, C_3, U_m, a_1, C_1, t_p, R_m, V_p) $  & $375$ & $863$ & $182$& $231$ & $632$ & $942$ & $224$ & $289$ & Lasso($\sigma$) \\ \hline
       \multicolumn{10}{|c|}{Method with the lowest MSE } \\ \hline \hline
        & Lasso & Lasso & Lasso/Union & Lasso & low-influence & Lasso & Lasso &  Lasso & \\ \hline
  \hline \hline
\end{tabular}
}
 \caption{The mean squared error (MSE) between forecast/smoothed and measured glucose. The machine-based methods, almost always selected the parameter set that achieved the MSE minimum, but for some individuals, certain hand-chosen parameters matter. }
   \label{table:howwhopickedwhatdid}
\end{table}

\begin{figure}
%\begin{adjustwidth}{-2.25in}{0in} % Comment out/remove adjustwidth environment if table fits in text column.
\centering
\includegraphics[scale=0.5]{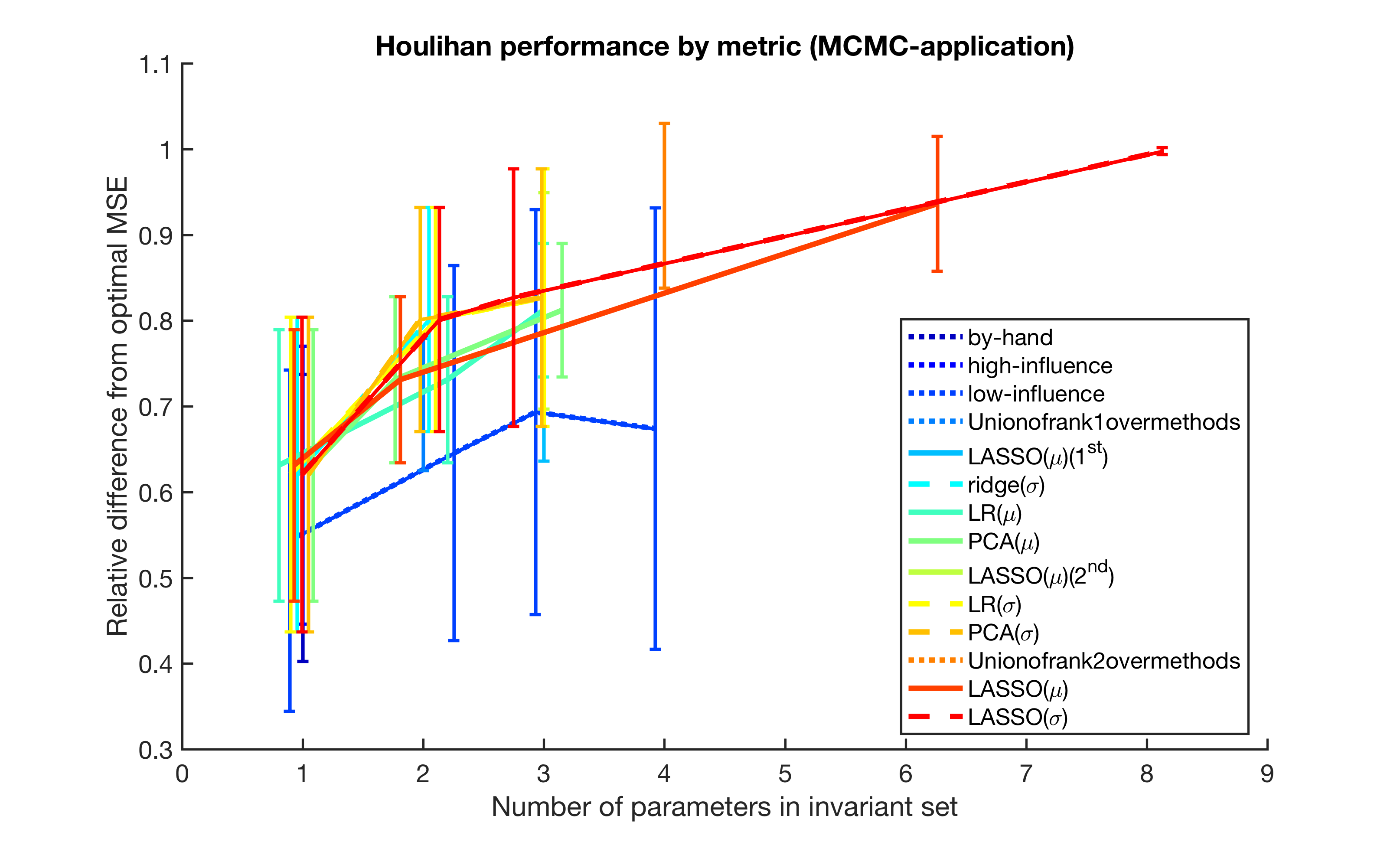}
\caption{The overall performance of each method in the smoothing setting.
The vertical axis indicates the $\%$-optimal MSE for a given method, averaged over the four patient data sets.
Note that methods are labeled as blue to red, where the minimally-performing methods are blue and the maximally-performing methods are red. The plots are estimated directly from the information in Table \ref{table:howwhopickedwhatdid}.}
\label{fig:compare_methods}
%\end{adjustwidth}
\end{figure}
%/Users/albers/Dropbox/high_throughput_identifability/src

To evaluate the effectiveness of the machine-selected parameters compared to low-influence parameters as characterized by their $\beta_i$'s, and the by-hand-selected parameters we used in our previous work, we compare the mean squared error (MSE) between the data and the forecasts for the various parameter combinations as shown in table \ref{table:howwhopickedwhatdid}. Fig \ref{fig:compare_methods} provides a visual summary of the results in table \ref{table:howwhopickedwhatdid}---the plots are calculated directly from table \ref{table:howwhopickedwhatdid}---for the MCMC smoothing setting, and demonstrates that all Houlihan-based parameter sets (of any size) noticeably out-performed the by-hand and low-influence parameter sets. Moreover, we see that most Houlihan-based methods achieve similar overall accuracy for parameter sets of cardinality $\leq 3$. In addition, Houlihan-based methods that selected parameter sets with 4 or more parameters achieved the best performance, and there is a general trend of improved fit with more parameters---this contrasts sharply with the by-hand parameter selections, whose performance tapered with more than 3 parameters (probably due to unforeseen issues of identifiability).

\begin{figure}
\centering
\includegraphics[scale=0.5]{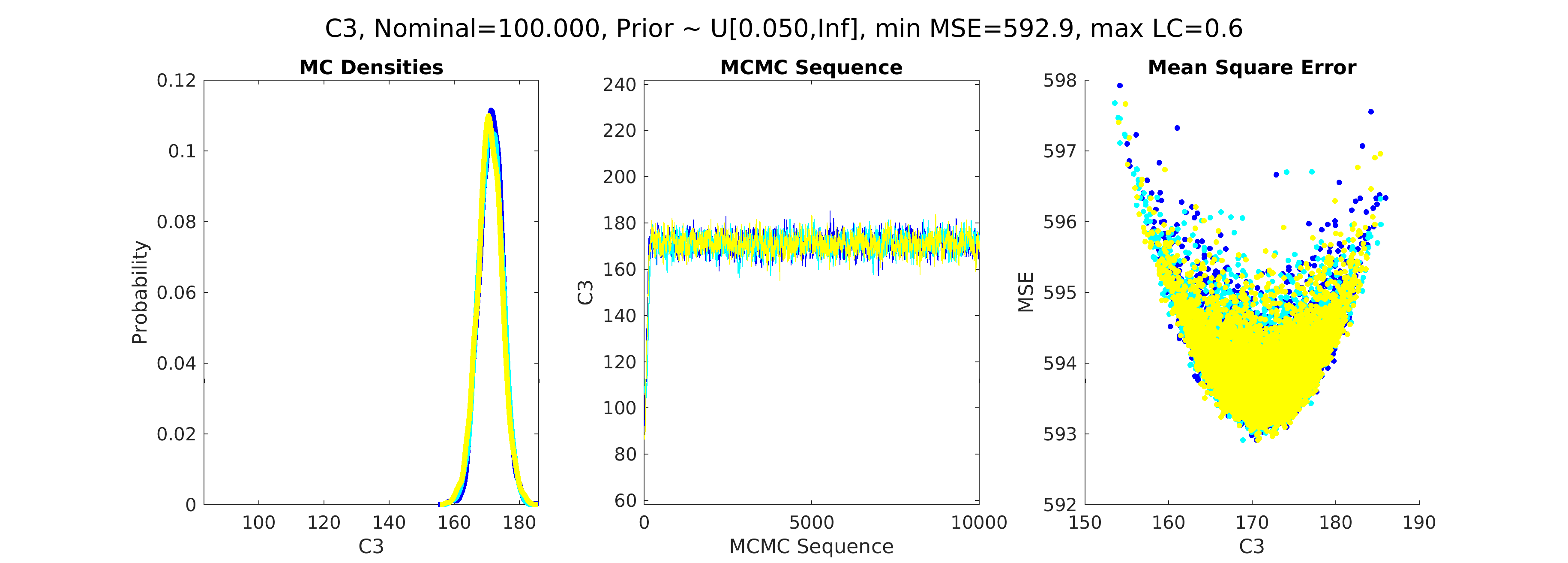}
\includegraphics[scale=0.5]{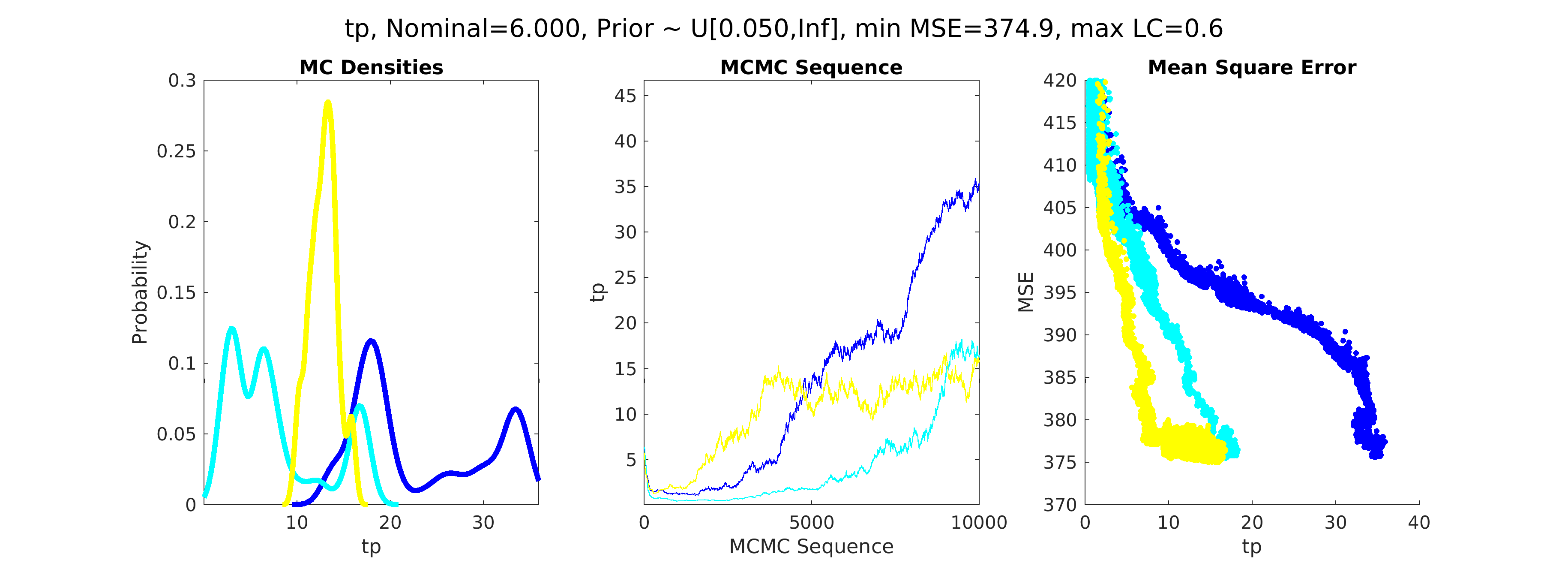}
\caption{The posterior densities, Markov chains, and MSE surfaces, for two parameters taken from two sets of parameters.  The top set of plots shows $C_3$ estimates for P1 where lasso is allowed to select two parameters with standard deviation set as the feature metric; $C_3$ converges well. The bottom set of plots show $t_p$ estimates for P1 for lasso-selected parameters at one standard error minimum---eight parameters are selected in this case---with standard deviation set as the feature metric; $t_p$ does not converge to a unique minimum but has a lower MSE than cases where the parameters are uniquely identified..}
\label{fig:mcmc_converence}
\end{figure}

In particular, lasso chose parameters with the lowest MSE between forecasts and measurements in $7$ of $8$ cases. In one case, taking the union over methods shared the same MSE with lasso.  And, in one case, the lowest MSE was observed with a pair of low-influence parameters. In this case it was the parameter-pair combination, $\alpha$ with $E$, that mattered.  This result implies that generally low influence parameters may, for some people, be physiologically important and explore particular pathophysiology necessary to synchronize to the individual. We also know that as the number of parameters increased to $3 \geq$, some of the MCMC parameter estimates with the lowest MSE found multiple, competing equilibria, were not unique, and sometimes did not fully converge. For example, Fig. \ref{fig:mcmc_converence} shows parameter estimates of two different parameters---one that converges and one that does not---for two parameter sets for P1 with standard deviation as the feature-metric.  When lasso-selected parameters are restricted to two parameters for P1, then both parameters, $R_g$ and $C_3$ converge producing a MSE of $600$; $C_3$ is shown in Fig. \ref{fig:mcmc_converence}. In contrast, lasso restricted to the one standard error minimum selects eight parameters, has a lower MSE of $375$ but at least one of the parameters, $t_p$, does not converge well as shown in Fig. \ref{fig:mcmc_converence}. This means that as  we increased the flexibility, we lowered the MSEs but possibly came at the expense of physiology or convergent parameter estimates.

%/Users/albers/Dropbox/high_throughput_identifability/src

\section{Discussion}

\textbf{Summary} Our most broad conclusion is that the machine-selected parameters work better than hand-selected parameters and that the Houlihan methods are a scalable method for selecting which parameters of a mechanistic model to estimate using DA methods. This means that stacking machine learning techniques on top of, or together with, DA is a helpful strategy, especially when models are complex and data are sparse, as in our glucose modeling example.

\textbf{Houlihan methods} We intuitively define the Houlihan method(s) as a collection of methods for selecting the most productive model parameters to estimate with machine learning techniques applied to simulated model output under parameter variation subject to a set of features, e.g., the mean of a state.

using machine learning to

\textbf{Feature metric selection matters}:  For all methods, the feature metric (mean or standard deviation) was the first-order driver of differences in parameter rank orderings.
% whether variation in the mean or the standard deviation was used to select parameters was the first order driver of differences in parameter selections.
This choice is highly problem-dependent. In some biomedical applications, sensitivity of the mean to parameter perturbation is not especially important for a good fit; e.g., there are physiologic systems where variation in the mean across people is small, but excursions, peaks, number of peaks, location of peaks, etc., may be a more important types of features to capture.

\textbf{The cutoff matters}: The cutoff for influence has a substantial impact on the ability to estimate parameters. For example, lasso-selected parameters usually minimized MSE, but the induced MSE and MCMC convergence were both sensitive to the influence cutoff. All the methods had this sensitivity, and estimating optimal cutoffs automatically would be beneficial.

\textbf{The selection method sometimes matters:} For the high-ranked parameter choices, the feature-metric was the primary difference between selected parameter sets. However, as the number of parameters included was increased, the methods diverged.  We suspect that as the complexity of feature metrics and ranking methods increases, e.g., using nonlinear regressions, there will be more sensitivity of the parameter selections to the methods.

\textbf{Physiology matters:} We know from carefully considering the convergence properties of the MCMC chains that some of the lowest MSEs for the runs with three or more parameters didn't converge well. Meaning, as we increased the flexibility, we lowered the MSE but possibly at the expense of physiologic fidelity or convergent parameter estimates. For pure forecasting applications this may or may not matter, but when we want the parameters to be meaningful, we need the parameter estimates to converge, not necessarily to \emph{a} unique set of parameters, but to distinctly different parameter estimates that can be treated as hypotheses. Another problem that can arise because of physiology is that different people with different physiology can be sensitive to different parameters. For example, the physiological feature that is important to personalize the model for a particular person may not be related to the properties captured by the feature metric, e.g., the mean, and in this circumstance parameters identified as low influence relative to the feature-metric will not be estimated. A potential example of this is P1, for whom estimating $\alpha$ and $E$ achieved the lowest MSE despite $E$ and $\alpha$ being low-influence parameters relative to both the mean and standard deviation.

\textbf{Effective parameter space exploration:} Abstractly, a mechanistic model is a parameterized family of functions whose parameters, depending on the model, have varying degrees of independence. From this perspective, the goal of the Houlihan methodology is to find a way to explore the maximal amount of the parameter space while minimizing the redundancy between parameters.
% that leads to non-uniqueness of optimal parameter estimates or identifiability failures.
The feature-metrics and the influence functions define which subsets of the parameter space are most useful to open for exploration, which in turn defines which dynamics can be explored. For example, focusing on variations of the mean may close off other dynamical features such as amplitude variations or any feature that is not uniquely defined by the variation of the mean. We do not yet have a good method for understanding  how a feature-metric may influence other, potentially valuable explorations. We acknowledge that understanding and quantifying how limited feature-metrics influence the effective parameter space of a model is an important, unexplored problem.

%DA contrasts non-mechanistic machine learning by having a limited model space to estimate
%
%We can treat the model as just a limited (incomplete non-dense, etc.) function space and not worry about properties such as iden
%
%
%
%But
%
% the person matters! the method is sensitive to the person. Effective parameter sapce
\textbf{Computational complexity:} We consider only the case here where we vary any one single parameter while leaving all other parameters fixed at their nominal values; this means that the dimension of the input for regression used to select the most useful parameters scales linearly in the number of parameters.  If we were to co-vary parameters, meaning if varied all parameters at once, depending on how one choose to partition the parameter space, the computational complexity would explode. In this way, the framework we present here does not solve the computational complexity problem of exploring parameter space. Instead, the results in this paper show that even by only considering feature-metric variation along one-dimensional subspaces of the full parameter space we can gain substantial insight into which parameters have the most impact on the features we are interested in approximating. Moreover, we can also see the limitations of this approach --- we do observe synergy between parameters where combinations of some low-influence parameters for some people can end up having a high influence on the model fit.

\textbf{Obvious extensions}: In this paper, we stack machine learning on top of DA, which has many potential extensions. Feature-metrics could be generalized to be multi-dimensional both over states and over types of feature-metrics. Feature selection methods could be developed or employed to select feature metrics. The estimates of influence could be calculated to include jointly varying parameters---this would be computationally expensive and would require computational innovation in high-dimensional settings, cf the computational complexity discussion above. Moreover, this problem is not necessarily a simple extension because the parameter spaces of mechanistic models are not likely to form a basis for the model space, in contrast to the parameters of the space of polynomials which do form a basis.  Of course this lack of a basis structure is part of the problem---parameters of mechanistic models and likely the physiology they represent are redundant, likely for biological reasons such as robustness. We use linear regression and PCA-based machine learning methods; it is likely that more sophisticated machine learning methods e.g., full elastic nets, support vector machines, deep learning, sparse machine learning (compressed sensing), Bayesian methods, model averaging and ensemble learning could all be used and would likely improve the parameter selections. Similarly, further stack of machine learning techniques on top of the Houlihan methods would likely be productive.  For example, greedy, Gibbs-sampling-like rotation between sets of parameters that are identifiable and explore different subsets of the parameter space could minimize both model errors and identifiability issues. And finally, feature-metrics could be made substantially more sophisticated, insightful and tailored to circumstance or physiologic knowledge, such as preserving power in certain frequency bands. More sophisticated feature metrics could also be used to gain insight into potentially meaningful constraints on parameters for use in operational DA.

\section{Conclusion}
We devised a methodology for rank-ordering parameters of a mechanistic model and using this rank-ordering to select an effective subset of parameters to estimate when projecting biomedical data onto the model via data assimilation.
This methodology specifically targets parameter sets that avoid issues of model identifiability and parameter-estimation convergence problems, improving forecasting and phenotyping performance of data assimilation methods that use mechanistic biological models.
Using machine learning to select parameters to estimate worked: the machine-chosen parameters reduced the mean-squared error between estimates and forecasts and data in nearly all cases by factors as large as three. These results imply that combining mechanistic and non-mechanistic machine learning could be a particularly productive direction of future research and could greatly aid in our ability to use computational machinery to both help deepen our physiologic understanding and help clinicians achieve more positive outcomes in clinical settings.

\appendix

\section{Ultradian model}
\label{app:ultradian}

The model is comprised of a set of six ordinary differential equations; the model is non-autonomous because it has an external, time-dependent driver, consumed nutrition. The six dimensional state space made up of three physiologic variables and a three stage filter.  The physiologic state variables are the glucose concentration $G$, the plasma insulin concentration $I_{p}$, and the interstitial insulin concentration $I_{i}$. The three stage filter $(h_1,h_2,h_3)$ which reflects the response of the plasma insulin to glucose levels \cite{sturis_91}. The model was designed to capture ultradian oscillations missing in previous models. The ordinary differential equations that define the model are \cite{keenerII}:
\begin{eqnarray}
\label{eq:model1}
\frac{dI_p}{dt} & = &  f_1(G)-E\bigl(\frac{I_{p}}{V_{p}}-\frac{I_i}{V_{i}}\bigr)-\frac{I_{p}}{t_{p}}\\
\frac{dI_i}{dt} & = & E\bigl(\frac{I_{p}}{V_{p}}-\frac{I_i}{V_{i}}\bigr)-\frac{I_{i}}{t_{i}}\\
\frac{dG}{dt} & = & f_4(h_3)+I_{G}(t)-f_2(G)-f_3(I_i)G\\
\frac{dh_1}{dt} & = & \frac{1}{t_d}\bigl(I_p-h_1\bigr) \\
\frac{dh_2}{dt} & = & \frac{1}{t_d}\bigl(h_1-h_2\bigr) \\
\frac{dh_3}{dt} & = & \frac{1}{t_d}\bigl(h_2-h_3\bigr)
\end{eqnarray}

%The \emph{major} parameters include: \emph{(i)} $E$, a rate constant for exchange of insulin between the plasma and remote compartments; \emph{(ii)} $I_G(t)$, the exogenous (externally driven) glucose delivery rate; \emph{(iii)} $t_p$, the time constant for plasma insulin degradation; \emph{(iv)} $t_i$, the time constant for the remote insulin degradation; \emph{(v)} $t_d$, the delay time between plasma insulin and glucose production; \emph{(vi)} $V_p$, the volume of insulin distribution in the plasma; \emph{(vii)} $V_i$, the volume of the remote insulin compartment; \emph{(viii)} $V_g$, the volume of the glucose space \cite{dyn_pheno} .

The state variables include physiologic processes that have been parameterized, including: $f_1(G)$ represents the rate of insulin production; $f_2(G)$ represents insulin-independent glucose utilization; $f_3(I_i)G$ represents insulin-dependent glucose utilization; $f_4(h_3)$ represents delayed insulin-dependent glucose utilization.  These functions are defined by:
\begin{eqnarray}
f_1(G) & = & \frac{R_m}{1+ \exp(\frac{-G}{V_g c_1} + a_1)} \\
f_2(G) & = & U_b(1-\exp(\frac{-G}{C_2V_g})) \\
f_3(I_i) & = & \frac{1}{C_3 V_g}( U_0 + \frac{U_m - U_0}{1 + (\kappa I_i)^{-\beta}}) \\
f_4(h_3) & = & \frac{R_g}{1 + \exp(\alpha (\frac{h_3}{C_5 V_p}  -1))} \\
\kappa & = & \frac{1}{C_4} (\frac{1}{V_i} - \frac{1}{E t_i})
\end{eqnarray}

The nutritional driver of the model $I_G(t)$ is defined over $N$ discrete nutrition events \cite{dyn_pheno}, where $k$ is the decay constant and event $j$ occurs at time $t_j$ with carbohydrate quantity $m_j$
\begin{equation}
I_G(t) = \sum^N_{j=1} \frac{m_j k}{60} \exp(k(t_j-t));  N=\# \{ t_j < t \}
\end{equation}

\begin{table}[!ht]
%\begin{adjustwidth}{-2.25in}{0in} % Comment out/remove adjustwidth environment if table fits in text column.
\centering
\caption{Full list of parameters for the ultradian glucose-insulin model \cite{keenerII}. Note that IIGU and IDGU denote insulin-independent glucose utilization and insulin-dependent glucose utilization, respectively.}
\begin{tabular}{|p{1.2cm}|p{2.7cm}|p{8cm}|}
\hline
\multicolumn{3}{|p{8cm}|}{\textbf{Ultradian model parameters}} \\ \hline
\Cline{2pt}{1-3} \hline
\Cline{2pt}{1-3}
Name & Nominal Value  & Meaning \\ \hline \hline
$V_p$  & $3$ l  & plasma volume  \\ \hline \hline
$V_i$  & $11$ l  & interstitial  volume \\ \hline \hline
$V_g$ & $10$ l  & glucose space \\ \hline \hline
$E$  & $0.2$ l min$^{-1}$ &   exchange rate for insulin between remote
and plasma compartments \\ \hline \hline
$t_p$  & $6$ min  & time constant for plasma insulin degradation (via
kidney and liver filtering) \\ \hline \hline
$t_i$  & $100$ min & time constant for remote insulin degradation (via
muscle and adipose tissue) \\ \hline \hline
$t_d$  & $12$ min  & delay between plasma insulin and glucose
production \\ \hline \hline
$k$  & $0.5$ min$^{-1}$  & rate of decayed appearance of ingested glucose \\ \hline \hline
$R_m$  & $209$ mU min$^{-1}$  & linear constant affecting insulin secretion  \\ \hline \hline
$a_1$  & $6.6$ & exponential constant affecting insulin secretion \\ \hline \hline
$C_1$  & $300$ mg l$^{-1}$ & exponential constant affecting insulin secretion \\ \hline \hline
$C_2$  & $144$ mg l$^{-1}$  & exponential constant affecting IIGU \\ \hline \hline
$C_3$  & $100$ mg l$^{-1}$  & linear constant affecting IDGU \\ \hline \hline
$C_4$  & $80$ mU l$^{-1}$ & factor affecting IDG \\ \hline \hline
$C_5$  & $26$ mU l$^{-1}$  & exponential constant affecting IDGU \\ \hline \hline
$U_b$  & $72$ mg min$^{-1}$  & linear constant affecting IIGU \\ \hline \hline
$U_0$  & $4$ mg min$^{-1}$ & linear constant affecting IDGU \\ \hline \hline
$U_m$  & $94$ mg min$^{-1}$  &  linear constant affecting IDGU \\ \hline \hline
$R_g$  & $180$ mg min$^{-1}$  & linear constant affecting IDGU \\ \hline \hline
$\alpha$  & $7.5$ & exponential constant affecting IDGU \\ \hline \hline
$\beta$  & $1.772$ & exponent affecting IDGU \\ \hline \hline
\end{tabular}
\label{table:model_parameters}
%\end{adjustwidth}
\end{table}

%% References


%\bibliographystyle{elsarticle-num}
\bibliography{master,dstexts,srb,structuralstability,partialhyperbolicity,neuralnetworks,bifurcationtheory}

\begin{thebibliography}{111}
\expandafter\ifx\csname natexlab\endcsname\relax\def\natexlab#1{#1}\fi
\expandafter\ifx\csname bibnamefont\endcsname\relax
  \def\bibnamefont#1{#1}\fi
\expandafter\ifx\csname bibfnamefont\endcsname\relax
  \def\bibfnamefont#1{#1}\fi
\expandafter\ifx\csname citenamefont\endcsname\relax
  \def\citenamefont#1{#1}\fi
\expandafter\ifx\csname url\endcsname\relax
  \def\url#1{\texttt{#1}}\fi
\expandafter\ifx\csname urlprefix\endcsname\relax\def\urlprefix{URL }\fi
\providecommand{\bibinfo}[2]{#2}
\providecommand{\eprint}[2][]{\url{#2}}

\bibitem[{\citenamefont{Albers et~al.}(2017)\citenamefont{Albers, Levine,
  Gluckman, Ginsberg, Hripcsak, and Mamykina}}]{da_glucose_forecast_t2d}
\bibinfo{author}{\bibfnamefont{D.}~\bibnamefont{Albers}},
  \bibinfo{author}{\bibfnamefont{M.}~\bibnamefont{Levine}},
  \bibinfo{author}{\bibfnamefont{B.}~\bibnamefont{Gluckman}},
  \bibinfo{author}{\bibfnamefont{H.}~\bibnamefont{Ginsberg}},
  \bibinfo{author}{\bibfnamefont{G.}~\bibnamefont{Hripcsak}}, \bibnamefont{and}
  \bibinfo{author}{\bibfnamefont{L.}~\bibnamefont{Mamykina}},
  \bibinfo{journal}{PloS Comp Bio} \textbf{\bibinfo{volume}{13}},
  \bibinfo{pages}{e1005232} (\bibinfo{year}{2017}).

\bibitem[{\citenamefont{Albers et~al.}(2012)\citenamefont{Albers, Hripcsak, and
  Schmidt}}]{pop_phys}
\bibinfo{author}{\bibfnamefont{D.}~\bibnamefont{Albers}},
  \bibinfo{author}{\bibfnamefont{G.}~\bibnamefont{Hripcsak}}, \bibnamefont{and}
  \bibinfo{author}{\bibfnamefont{M.}~\bibnamefont{Schmidt}},
  \bibinfo{journal}{PLoS One} \textbf{\bibinfo{volume}{7}},
  \bibinfo{pages}{e480058} (\bibinfo{year}{2012}).

\bibitem[{\citenamefont{Xu et~al.}(2016)\citenamefont{Xu, Xu, and
  Saria}}]{pmlr-v56-Xu16}
\bibinfo{author}{\bibfnamefont{Y.}~\bibnamefont{Xu}},
  \bibinfo{author}{\bibfnamefont{Y.}~\bibnamefont{Xu}}, \bibnamefont{and}
  \bibinfo{author}{\bibfnamefont{S.}~\bibnamefont{Saria}}, in
  \emph{\bibinfo{booktitle}{Proceedings of the 1st Machine Learning for
  Healthcare Conference}}, edited by
  \bibinfo{editor}{\bibfnamefont{F.}~\bibnamefont{Doshi-Velez}},
  \bibinfo{editor}{\bibfnamefont{J.}~\bibnamefont{Fackler}},
  \bibinfo{editor}{\bibfnamefont{D.}~\bibnamefont{Kale}},
  \bibinfo{editor}{\bibfnamefont{B.}~\bibnamefont{Wallace}}, \bibnamefont{and}
  \bibinfo{editor}{\bibfnamefont{J.}~\bibnamefont{Weins}}
  (\bibinfo{publisher}{PMLR}, \bibinfo{address}{Northeastern University,
  Boston, MA, USA}, \bibinfo{year}{2016}), vol.~\bibinfo{volume}{56} of
  \emph{\bibinfo{series}{Proceedings of Machine Learning Research}}, pp.
  \bibinfo{pages}{282--300},
  \urlprefix\url{http://proceedings.mlr.press/v56/Xu16.html}.

\bibitem[{\citenamefont{Albers et~al.}(2014)\citenamefont{Albers, Elhadad,
  Tabak, Perotte, and Hripcsak}}]{dyn_pheno}
\bibinfo{author}{\bibfnamefont{D.}~\bibnamefont{Albers}},
  \bibinfo{author}{\bibfnamefont{N.}~\bibnamefont{Elhadad}},
  \bibinfo{author}{\bibfnamefont{E.}~\bibnamefont{Tabak}},
  \bibinfo{author}{\bibfnamefont{A.}~\bibnamefont{Perotte}}, \bibnamefont{and}
  \bibinfo{author}{\bibfnamefont{G.}~\bibnamefont{Hripcsak}},
  \bibinfo{journal}{PLoS One} \textbf{\bibinfo{volume}{6}},
  \bibinfo{pages}{e96443} (\bibinfo{year}{2014}).

\bibitem[{\citenamefont{Hripcsak and Albers}(2017)}]{high_fide_pheno}
\bibinfo{author}{\bibfnamefont{G.}~\bibnamefont{Hripcsak}} \bibnamefont{and}
  \bibinfo{author}{\bibfnamefont{D.~J.} \bibnamefont{Albers}},
  \bibinfo{journal}{Journal of the American Medical Informatics Association} p.
  \bibinfo{pages}{ocx110} (\bibinfo{year}{2017}).

\bibitem[{\citenamefont{Pathak et~al.}(2013)\citenamefont{Pathak, Kho, and
  Denny}}]{emerge3}
\bibinfo{author}{\bibfnamefont{J.}~\bibnamefont{Pathak}},
  \bibinfo{author}{\bibfnamefont{A.}~\bibnamefont{Kho}}, \bibnamefont{and}
  \bibinfo{author}{\bibfnamefont{J.}~\bibnamefont{Denny}}, \bibinfo{journal}{J
  Am Med Inform Assoc.} \textbf{\bibinfo{volume}{20}}, \bibinfo{pages}{e206}
  (\bibinfo{year}{2013}).

\bibitem[{\citenamefont{Pivovarov et~al.}(2015)\citenamefont{Pivovarov,
  Perotte, Grave, Angiolillo, Wiggins, and Elhadad}}]{phenome_model}
\bibinfo{author}{\bibfnamefont{R.}~\bibnamefont{Pivovarov}},
  \bibinfo{author}{\bibfnamefont{A.}~\bibnamefont{Perotte}},
  \bibinfo{author}{\bibfnamefont{E.}~\bibnamefont{Grave}},
  \bibinfo{author}{\bibfnamefont{J.}~\bibnamefont{Angiolillo}},
  \bibinfo{author}{\bibfnamefont{C.}~\bibnamefont{Wiggins}}, \bibnamefont{and}
  \bibinfo{author}{\bibfnamefont{N.}~\bibnamefont{Elhadad}},
  \bibinfo{journal}{Journal of Biomedical Informatics}  (\bibinfo{year}{2015}).

\bibitem[{\citenamefont{Halpern et~al.}(2016)\citenamefont{Halpern, Choi,
  Horng, and Sontag}}]{anchor_phenome_jamia}
\bibinfo{author}{\bibfnamefont{Y.}~\bibnamefont{Halpern}},
  \bibinfo{author}{\bibfnamefont{Y.}~\bibnamefont{Choi}},
  \bibinfo{author}{\bibfnamefont{S.}~\bibnamefont{Horng}}, \bibnamefont{and}
  \bibinfo{author}{\bibfnamefont{D.}~\bibnamefont{Sontag}},
  \bibinfo{journal}{JAMIA}  (\bibinfo{year}{2016}).

\bibitem[{\citenamefont{Rajkomar et~al.}(2018)\citenamefont{Rajkomar, Oren,
  Chen, Dai, Hajaj, Liu, Liu, Sun, Sundberg, Yee et~al.}}]{google_deep_pheno}
\bibinfo{author}{\bibfnamefont{A.}~\bibnamefont{Rajkomar}},
  \bibinfo{author}{\bibfnamefont{E.}~\bibnamefont{Oren}},
  \bibinfo{author}{\bibfnamefont{K.}~\bibnamefont{Chen}},
  \bibinfo{author}{\bibfnamefont{A.~M.} \bibnamefont{Dai}},
  \bibinfo{author}{\bibfnamefont{N.}~\bibnamefont{Hajaj}},
  \bibinfo{author}{\bibfnamefont{P.~J.} \bibnamefont{Liu}},
  \bibinfo{author}{\bibfnamefont{X.}~\bibnamefont{Liu}},
  \bibinfo{author}{\bibfnamefont{M.}~\bibnamefont{Sun}},
  \bibinfo{author}{\bibfnamefont{P.}~\bibnamefont{Sundberg}},
  \bibinfo{author}{\bibfnamefont{H.}~\bibnamefont{Yee}}, \bibnamefont{et~al.},
  \bibinfo{journal}{CoRR} \textbf{\bibinfo{volume}{abs/1801.07860}}
  (\bibinfo{year}{2018}), \eprint{1801.07860},
  \urlprefix\url{http://arxiv.org/abs/1801.07860}.

\bibitem[{\citenamefont{Westwick and Kearney}(2003)}]{physio_uncertainty_book}
\bibinfo{author}{\bibfnamefont{D.~T.} \bibnamefont{Westwick}} \bibnamefont{and}
  \bibinfo{author}{\bibfnamefont{R.~E.} \bibnamefont{Kearney}},
  \emph{\bibinfo{title}{Identification of nonlinear physiological systems}}
  (\bibinfo{publisher}{IEEE Engineering in Medicine and Biology},
  \bibinfo{year}{2003}).

\bibitem[{\citenamefont{Ljung}(1987)}]{system_id_ljun}
\bibinfo{author}{\bibfnamefont{L.}~\bibnamefont{Ljung}},
  \emph{\bibinfo{title}{System Identification}} (\bibinfo{publisher}{Prentice
  Hall}, \bibinfo{year}{1987}).

\bibitem[{\citenamefont{Schoukens et~al.}(2016)\citenamefont{Schoukens, Vaes,
  and Pintelon}}]{linear_id_nonlinear_setting}
\bibinfo{author}{\bibfnamefont{J.}~\bibnamefont{Schoukens}},
  \bibinfo{author}{\bibfnamefont{M.}~\bibnamefont{Vaes}}, \bibnamefont{and}
  \bibinfo{author}{\bibfnamefont{R.}~\bibnamefont{Pintelon}},
  \bibinfo{journal}{IEEE Control Systems} pp. \bibinfo{pages}{38--69}
  (\bibinfo{year}{2016}).

\bibitem[{\citenamefont{Levine et~al.}(2016)\citenamefont{Levine, Albers, and
  Hripcsak}}]{amia_matt_george_granger_like}
\bibinfo{author}{\bibfnamefont{M.}~\bibnamefont{Levine}},
  \bibinfo{author}{\bibfnamefont{D.}~\bibnamefont{Albers}}, \bibnamefont{and}
  \bibinfo{author}{\bibfnamefont{G.}~\bibnamefont{Hripcsak}}, in
  \emph{\bibinfo{booktitle}{Annual Symposium Proceedings}}
  (\bibinfo{organization}{AMIA}, \bibinfo{year}{2016}).

\bibitem[{\citenamefont{Levine et~al.}(2018)\citenamefont{Levine, Albers, and
  Hripcsak}}]{full_lagged_regression_matt}
\bibinfo{author}{\bibfnamefont{M.}~\bibnamefont{Levine}},
  \bibinfo{author}{\bibfnamefont{D.}~\bibnamefont{Albers}}, \bibnamefont{and}
  \bibinfo{author}{\bibfnamefont{G.}~\bibnamefont{Hripcsak}}
  (\bibinfo{year}{2018}), \bibinfo{note}{arXiv:1801.08929}.

\bibitem[{\citenamefont{Hripcsak et~al.}(2011)\citenamefont{Hripcsak, Albers,
  and Perotte}}]{george_lagged_correlation_jamia}
\bibinfo{author}{\bibfnamefont{G.}~\bibnamefont{Hripcsak}},
  \bibinfo{author}{\bibfnamefont{D.}~\bibnamefont{Albers}}, \bibnamefont{and}
  \bibinfo{author}{\bibfnamefont{A.}~\bibnamefont{Perotte}},
  \bibinfo{journal}{JAMIA} \textbf{\bibinfo{volume}{18}}, \bibinfo{pages}{109}
  (\bibinfo{year}{2011}).

\bibitem[{\citenamefont{Hripcsak and Albers}(2013)}]{jamia_hcpmodel13}
\bibinfo{author}{\bibfnamefont{G.}~\bibnamefont{Hripcsak}} \bibnamefont{and}
  \bibinfo{author}{\bibfnamefont{D.}~\bibnamefont{Albers}},
  \bibinfo{journal}{JAMIA} \textbf{\bibinfo{volume}{0}}, \bibinfo{pages}{1}
  (\bibinfo{year}{2013}).

\bibitem[{\citenamefont{Perotte et~al.}(2015)\citenamefont{Perotte, Ranganath,
  Hirsch, Blei, and Elhadad}}]{deep_survival_I}
\bibinfo{author}{\bibfnamefont{A.}~\bibnamefont{Perotte}},
  \bibinfo{author}{\bibfnamefont{R.}~\bibnamefont{Ranganath}},
  \bibinfo{author}{\bibfnamefont{J.}~\bibnamefont{Hirsch}},
  \bibinfo{author}{\bibfnamefont{D.}~\bibnamefont{Blei}}, \bibnamefont{and}
  \bibinfo{author}{\bibfnamefont{N.}~\bibnamefont{Elhadad}},
  \bibinfo{journal}{JAMIA} \textbf{\bibinfo{volume}{22}}, \bibinfo{pages}{872}
  (\bibinfo{year}{2015}).

\bibitem[{\citenamefont{Ranganath et~al.}(2016)\citenamefont{Ranganath,
  Perotte, Elhadad, and Blei}}]{deep_survival_II}
\bibinfo{author}{\bibfnamefont{R.}~\bibnamefont{Ranganath}},
  \bibinfo{author}{\bibfnamefont{A.}~\bibnamefont{Perotte}},
  \bibinfo{author}{\bibfnamefont{N.}~\bibnamefont{Elhadad}}, \bibnamefont{and}
  \bibinfo{author}{\bibfnamefont{D.}~\bibnamefont{Blei}}, in
  \emph{\bibinfo{booktitle}{Proceedings of Machine Learning for Healthcare}}
  (\bibinfo{year}{2016}), vol.~\bibinfo{volume}{56}.

\bibitem[{\citenamefont{Lasko et~al.}(2013)\citenamefont{Lasko, Denny, and
  Levy}}]{lasko_plos}
\bibinfo{author}{\bibfnamefont{T.}~\bibnamefont{Lasko}},
  \bibinfo{author}{\bibfnamefont{J.}~\bibnamefont{Denny}}, \bibnamefont{and}
  \bibinfo{author}{\bibfnamefont{M.}~\bibnamefont{Levy}},
  \bibinfo{journal}{PLOS One}  (\bibinfo{year}{2013}).

\bibitem[{\citenamefont{Hornik et~al.}(1989)\citenamefont{Hornik, Stinchocombe,
  and White}}]{hor}
\bibinfo{author}{\bibfnamefont{K.}~\bibnamefont{Hornik}},
  \bibinfo{author}{\bibfnamefont{M.}~\bibnamefont{Stinchocombe}},
  \bibnamefont{and} \bibinfo{author}{\bibfnamefont{H.}~\bibnamefont{White}},
  \bibinfo{journal}{Neural Networks} \textbf{\bibinfo{volume}{2}},
  \bibinfo{pages}{359} (\bibinfo{year}{1989}).

\bibitem[{\citenamefont{Hornik et~al.}(1990)\citenamefont{Hornik, Stinchocombe,
  and White}}]{hor2}
\bibinfo{author}{\bibfnamefont{K.}~\bibnamefont{Hornik}},
  \bibinfo{author}{\bibfnamefont{M.}~\bibnamefont{Stinchocombe}},
  \bibnamefont{and} \bibinfo{author}{\bibfnamefont{H.}~\bibnamefont{White}},
  \bibinfo{journal}{Neural Networks} \textbf{\bibinfo{volume}{3}},
  \bibinfo{pages}{551} (\bibinfo{year}{1990}).

\bibitem[{\citenamefont{Hripcsak and Albers}(2012)}]{jamia_phys_ehr}
\bibinfo{author}{\bibfnamefont{G.}~\bibnamefont{Hripcsak}} \bibnamefont{and}
  \bibinfo{author}{\bibfnamefont{D.}~\bibnamefont{Albers}},
  \bibinfo{journal}{JAMIA} \textbf{\bibinfo{volume}{10}}, \bibinfo{pages}{1}
  (\bibinfo{year}{2012}).

\bibitem[{\citenamefont{Pivovarov et~al.}(2014)\citenamefont{Pivovarov, Albers,
  Sepulveda, and Elhadad}}]{measurement_dynamics_rimma}
\bibinfo{author}{\bibfnamefont{R.}~\bibnamefont{Pivovarov}},
  \bibinfo{author}{\bibfnamefont{D.}~\bibnamefont{Albers}},
  \bibinfo{author}{\bibfnamefont{J.}~\bibnamefont{Sepulveda}},
  \bibnamefont{and} \bibinfo{author}{\bibfnamefont{N.}~\bibnamefont{Elhadad}},
  \bibinfo{journal}{Journal of Biomedical Informatics}  (\bibinfo{year}{2014}).

\bibitem[{\citenamefont{Keener and Sneyd}(2008)}]{keenerII}
\bibinfo{author}{\bibfnamefont{J.}~\bibnamefont{Keener}} \bibnamefont{and}
  \bibinfo{author}{\bibfnamefont{J.}~\bibnamefont{Sneyd}},
  \emph{\bibinfo{title}{Mathematical physiology II: Systems physiology}}
  (\bibinfo{publisher}{Springer}, \bibinfo{year}{2008}).

\bibitem[{\citenamefont{Brin and Stuck}(2004)}]{brin_ds_book}
\bibinfo{author}{\bibfnamefont{M.}~\bibnamefont{Brin}} \bibnamefont{and}
  \bibinfo{author}{\bibfnamefont{G.}~\bibnamefont{Stuck}},
  \emph{\bibinfo{title}{Introduction to Dynamical Systems}}
  (\bibinfo{publisher}{Cambridge University Press}, \bibinfo{year}{2004}).

\bibitem[{\citenamefont{Guckenheimer and Holmes}(1983)}]{guck}
\bibinfo{author}{\bibfnamefont{J.}~\bibnamefont{Guckenheimer}}
  \bibnamefont{and} \bibinfo{author}{\bibfnamefont{P.}~\bibnamefont{Holmes}},
  \emph{\bibinfo{title}{Nonlinear Oscillaions, Dynamical Systems, and
  Bifurcations of Vector Fields}} (\bibinfo{publisher}{Springer-Verlag, New
  York}, \bibinfo{year}{1983}).

\bibitem[{\citenamefont{Arrowsmith and Place}(1990)}]{arrowsmithandplace}
\bibinfo{author}{\bibfnamefont{D.~K.} \bibnamefont{Arrowsmith}}
  \bibnamefont{and} \bibinfo{author}{\bibfnamefont{C.~M.} \bibnamefont{Place}},
  \emph{\bibinfo{title}{An introduction to dynamical systems}}
  (\bibinfo{publisher}{Cambridge {U}niversity {P}ress}, \bibinfo{year}{1990}).

\bibitem[{\citenamefont{Arnold}(1992)}]{arnoldode}
\bibinfo{author}{\bibfnamefont{V.~I.} \bibnamefont{Arnold}},
  \emph{\bibinfo{title}{Ordinary differential equations}}
  (\bibinfo{publisher}{Springer-Verlag}, \bibinfo{year}{1992}).

\bibitem[{\citenamefont{Arnold}(1983)}]{arnoldgeo}
\bibinfo{author}{\bibfnamefont{V.}~\bibnamefont{Arnold}},
  \emph{\bibinfo{title}{Geometric methods in the theory of ordinary
  differential equations.}}, Grundlehren de mathematischen {W}issenschaften
  (\bibinfo{publisher}{Springer-Verlag}, \bibinfo{year}{1983}),
  \bibinfo{edition}{$2^{nd}$} ed.

\bibitem[{\citenamefont{Albers et~al.}(2018)\citenamefont{Albers, Levine,
  Gluckman, Hripcsak, Mamykina, and Stuart}}]{daJAMIA}
\bibinfo{author}{\bibfnamefont{D.}~\bibnamefont{Albers}},
  \bibinfo{author}{\bibfnamefont{L.}~\bibnamefont{Levine}},
  \bibinfo{author}{\bibfnamefont{B.}~\bibnamefont{Gluckman}},
  \bibinfo{author}{\bibfnamefont{G.}~\bibnamefont{Hripcsak}},
  \bibinfo{author}{\bibfnamefont{L.}~\bibnamefont{Mamykina}}, \bibnamefont{and}
  \bibinfo{author}{\bibfnamefont{A.}~\bibnamefont{Stuart}}
  (\bibinfo{year}{2018}), \bibinfo{note}{in revision}.

\bibitem[{\citenamefont{Jazwinski}(1998)}]{filtering_jaz}
\bibinfo{author}{\bibfnamefont{A.}~\bibnamefont{Jazwinski}},
  \emph{\bibinfo{title}{Stochastic processes and Filtering Theory}}
  (\bibinfo{publisher}{Dover}, \bibinfo{year}{1998}).

\bibitem[{\citenamefont{Lorenc}(1988)}]{data_assimilation_I}
\bibinfo{author}{\bibfnamefont{A.}~\bibnamefont{Lorenc}}, \bibinfo{journal}{Q.
  J. R. Meterol. Soc.} \textbf{\bibinfo{volume}{112}}, \bibinfo{pages}{1177}
  (\bibinfo{year}{1988}).

\bibitem[{\citenamefont{Law et~al.}(2015)\citenamefont{Law, Stuart, and
  Zygalakis}}]{stuart_da}
\bibinfo{author}{\bibfnamefont{K.}~\bibnamefont{Law}},
  \bibinfo{author}{\bibfnamefont{A.}~\bibnamefont{Stuart}}, \bibnamefont{and}
  \bibinfo{author}{\bibfnamefont{K.}~\bibnamefont{Zygalakis}},
  \emph{\bibinfo{title}{Data assimilation}} (\bibinfo{publisher}{Springer},
  \bibinfo{year}{2015}).

\bibitem[{\citenamefont{Ash et~al.}(2016)\citenamefont{Ash, Bocquet, and
  Nodet}}]{french_DA_asch}
\bibinfo{author}{\bibfnamefont{M.}~\bibnamefont{Ash}},
  \bibinfo{author}{\bibfnamefont{M.}~\bibnamefont{Bocquet}}, \bibnamefont{and}
  \bibinfo{author}{\bibfnamefont{M.}~\bibnamefont{Nodet}},
  \emph{\bibinfo{title}{Data assimilation: methods, algorithms and
  applications}} (\bibinfo{publisher}{SIAM}, \bibinfo{year}{2016}).

\bibitem[{\citenamefont{Reich and Cotter}(2015)}]{da_sebastian}
\bibinfo{author}{\bibfnamefont{S.}~\bibnamefont{Reich}} \bibnamefont{and}
  \bibinfo{author}{\bibfnamefont{C.}~\bibnamefont{Cotter}},
  \emph{\bibinfo{title}{Probabilistic forecasting and Bayesian data
  assimilation}} (\bibinfo{publisher}{Cambridge University Press},
  \bibinfo{year}{2015}).

\bibitem[{\citenamefont{Haug}(2012)}]{bayesian_estimation_tracking}
\bibinfo{author}{\bibfnamefont{A.}~\bibnamefont{Haug}},
  \emph{\bibinfo{title}{Baysian estimation and tracking}}
  (\bibinfo{publisher}{Wiley}, \bibinfo{year}{2012}).

\bibitem[{\citenamefont{Ristic et~al.}(2004)\citenamefont{Ristic, Arulampalam,
  and Gordon}}]{beyond_kalman}
\bibinfo{author}{\bibfnamefont{B.}~\bibnamefont{Ristic}},
  \bibinfo{author}{\bibfnamefont{S.}~\bibnamefont{Arulampalam}},
  \bibnamefont{and} \bibinfo{author}{\bibfnamefont{N.}~\bibnamefont{Gordon}},
  \emph{\bibinfo{title}{Beyond the {K}alman filter: particle filters for
  tracking and applications}} (\bibinfo{publisher}{Artech house},
  \bibinfo{year}{2004}).

\bibitem[{\citenamefont{Candy}(2009)}]{baysian_signal_processing}
\bibinfo{author}{\bibfnamefont{J.}~\bibnamefont{Candy}},
  \emph{\bibinfo{title}{Bayesian signal processing: classical, modern, and
  particle filtering methods}} (\bibinfo{publisher}{Wiley},
  \bibinfo{year}{2009}).

\bibitem[{\citenamefont{Evensen}(2009)}]{DA_aos_evensen}
\bibinfo{author}{\bibfnamefont{G.}~\bibnamefont{Evensen}},
  \emph{\bibinfo{title}{Data assimilation, the ensemble kalman filter}}
  (\bibinfo{publisher}{Springer}, \bibinfo{year}{2009}).

\bibitem[{\citenamefont{Evensen}(2003)}]{evensen_enkf_early}
\bibinfo{author}{\bibfnamefont{G.}~\bibnamefont{Evensen}},
  \bibinfo{journal}{Ocean Dynamics}  (\bibinfo{year}{2003}).

\bibitem[{\citenamefont{Stuart}(2010)}]{andrew_inverse_problems}
\bibinfo{author}{\bibfnamefont{A.}~\bibnamefont{Stuart}},
  \bibinfo{journal}{Acta Numerica} \textbf{\bibinfo{volume}{19}},
  \bibinfo{pages}{451} (\bibinfo{year}{2010}).

\bibitem[{\citenamefont{Zenker et~al.}(2007)\citenamefont{Zenker, Rubin, and
  Clermont}}]{clermont_inverse_problems}
\bibinfo{author}{\bibfnamefont{S.}~\bibnamefont{Zenker}},
  \bibinfo{author}{\bibfnamefont{J.}~\bibnamefont{Rubin}}, \bibnamefont{and}
  \bibinfo{author}{\bibfnamefont{G.}~\bibnamefont{Clermont}},
  \bibinfo{journal}{PLoS Comput Biol} \textbf{\bibinfo{volume}{3}}
  (\bibinfo{year}{2007}).

\bibitem[{\citenamefont{Banks et~al.}(2014)\citenamefont{Banks, Hu, , and
  Thompson}}]{inverse_book_banks}
\bibinfo{author}{\bibfnamefont{H.~T.} \bibnamefont{Banks}},
  \bibinfo{author}{\bibfnamefont{S.}~\bibnamefont{Hu}}, , \bibnamefont{and}
  \bibinfo{author}{\bibfnamefont{W.~C.} \bibnamefont{Thompson}},
  \emph{\bibinfo{title}{Modeling and inverse problems in the presence of
  uncertainty}} (\bibinfo{publisher}{CRC Press}, \bibinfo{year}{2014}).

\bibitem[{\citenamefont{Hirata et~al.}(2010{\natexlab{a}})\citenamefont{Hirata,
  Bruchovsky, and Aihara}}]{hirata_prostate_model}
\bibinfo{author}{\bibfnamefont{Y.}~\bibnamefont{Hirata}},
  \bibinfo{author}{\bibfnamefont{N.}~\bibnamefont{Bruchovsky}},
  \bibnamefont{and} \bibinfo{author}{\bibfnamefont{K.}~\bibnamefont{Aihara}},
  \bibinfo{journal}{Journal of Theoretical Biology,}
  \textbf{\bibinfo{volume}{264}}, \bibinfo{pages}{517}
  (\bibinfo{year}{2010}{\natexlab{a}}).

\bibitem[{\citenamefont{Hirata et~al.}(2010{\natexlab{b}})\citenamefont{Hirata,
  de~Bernardo, Bruchovsky, and Aihara}}]{hirata_prostate_cancer_chaos}
\bibinfo{author}{\bibfnamefont{Y.}~\bibnamefont{Hirata}},
  \bibinfo{author}{\bibfnamefont{M.}~\bibnamefont{de~Bernardo}},
  \bibinfo{author}{\bibfnamefont{N.}~\bibnamefont{Bruchovsky}},
  \bibnamefont{and} \bibinfo{author}{\bibfnamefont{K.}~\bibnamefont{Aihara}},
  \bibinfo{journal}{CHAOS} \textbf{\bibinfo{volume}{20}},
  \bibinfo{pages}{0451251} (\bibinfo{year}{2010}{\natexlab{b}}).

\bibitem[{\citenamefont{Schiff}(2011)}]{schiff_neuro_control_theory}
\bibinfo{author}{\bibfnamefont{S.}~\bibnamefont{Schiff}},
  \emph{\bibinfo{title}{Neural control engineering: The emerging intersection
  between control theory and neuroscience}} (\bibinfo{publisher}{MIT Press},
  \bibinfo{year}{2011}).

\bibitem[{\citenamefont{Dukic et~al.}(2012)\citenamefont{Dukic, Lopes, and
  Polson}}]{vanja_sir}
\bibinfo{author}{\bibfnamefont{V.}~\bibnamefont{Dukic}},
  \bibinfo{author}{\bibfnamefont{H.}~\bibnamefont{Lopes}}, \bibnamefont{and}
  \bibinfo{author}{\bibfnamefont{N.}~\bibnamefont{Polson}},
  \bibinfo{journal}{Journal of the American Statistical Association}
  \textbf{\bibinfo{volume}{107}}, \bibinfo{pages}{1410} (\bibinfo{year}{2012}).

\bibitem[{\citenamefont{Miao et~al.}(2011)\citenamefont{Miao, Xia, Perelson,
  and Wu}}]{SIAM_identifiability_bio_modeling}
\bibinfo{author}{\bibfnamefont{H.}~\bibnamefont{Miao}},
  \bibinfo{author}{\bibfnamefont{X.}~\bibnamefont{Xia}},
  \bibinfo{author}{\bibfnamefont{A.}~\bibnamefont{Perelson}}, \bibnamefont{and}
  \bibinfo{author}{\bibfnamefont{H.}~\bibnamefont{Wu}}, \bibinfo{journal}{SIAM
  Review} \textbf{\bibinfo{volume}{53}}, \bibinfo{pages}{3}
  (\bibinfo{year}{2011}).

\bibitem[{\citenamefont{Chee and Fernando}(2007)}]{closed_loop_glucose_control}
\bibinfo{author}{\bibfnamefont{F.}~\bibnamefont{Chee}} \bibnamefont{and}
  \bibinfo{author}{\bibfnamefont{T.}~\bibnamefont{Fernando}},
  \emph{\bibinfo{title}{Closed-loop control of blood glucose}}
  (\bibinfo{publisher}{Springer}, \bibinfo{year}{2007}).

\bibitem[{\citenamefont{P et~al.}(2003)\citenamefont{P, Orsini, and
  MMBenedetti}}]{artificial_pancreas_I}
\bibinfo{author}{\bibfnamefont{P.~B.} \bibnamefont{P}},
  \bibinfo{author}{\bibfnamefont{M.}~\bibnamefont{Orsini}}, \bibnamefont{and}
  \bibinfo{author}{\bibnamefont{MMBenedetti}}, \bibinfo{journal}{Artif Cells
  Blood Substit Immobil Biotechnol.} \textbf{\bibinfo{volume}{2}},
  \bibinfo{pages}{127} (\bibinfo{year}{2003}).

\bibitem[{\citenamefont{Fabietti et~al.}(2007)\citenamefont{Fabietti, Canonico,
  Orsini-Federici, Sarti, and Massi-Benedetti}}]{artificial_pancreas_II}
\bibinfo{author}{\bibfnamefont{P.}~\bibnamefont{Fabietti}},
  \bibinfo{author}{\bibfnamefont{V.}~\bibnamefont{Canonico}},
  \bibinfo{author}{\bibfnamefont{M.}~\bibnamefont{Orsini-Federici}},
  \bibinfo{author}{\bibfnamefont{E.}~\bibnamefont{Sarti}}, \bibnamefont{and}
  \bibinfo{author}{\bibfnamefont{M.}~\bibnamefont{Massi-Benedetti}},
  \bibinfo{journal}{Diabetes Technol Ther.} \textbf{\bibinfo{volume}{4}},
  \bibinfo{pages}{327} (\bibinfo{year}{2007}).

\bibitem[{\citenamefont{Kovatchev et~al.}(2009)\citenamefont{Kovatchev, Breton,
  Man, and Cobelli}}]{in_silico_glucose_model_fda}
\bibinfo{author}{\bibfnamefont{B.}~\bibnamefont{Kovatchev}},
  \bibinfo{author}{\bibfnamefont{M.}~\bibnamefont{Breton}},
  \bibinfo{author}{\bibfnamefont{C.}~\bibnamefont{Man}}, \bibnamefont{and}
  \bibinfo{author}{\bibfnamefont{C.}~\bibnamefont{Cobelli}},
  \bibinfo{journal}{J Diabetes Sci Technol} \textbf{\bibinfo{volume}{3}},
  \bibinfo{pages}{44} (\bibinfo{year}{2009}).

\bibitem[{\citenamefont{Coutinho et~al.}(2003)\citenamefont{Coutinho, Azevedo,
  Burattini, Lopenz, and Massad}}]{rubella_model}
\bibinfo{author}{\bibfnamefont{M.~A. F. A.~B.} \bibnamefont{Coutinho}},
  \bibinfo{author}{\bibfnamefont{R.~S.} \bibnamefont{Azevedo}},
  \bibinfo{author}{\bibfnamefont{M.~N.} \bibnamefont{Burattini}},
  \bibinfo{author}{\bibfnamefont{L.~F.} \bibnamefont{Lopenz}},
  \bibnamefont{and} \bibinfo{author}{\bibfnamefont{E.}~\bibnamefont{Massad}},
  \bibinfo{journal}{Phys. Rev. E} \textbf{\bibinfo{volume}{67}},
  \bibinfo{pages}{051907} (\bibinfo{year}{2003}).

\bibitem[{\citenamefont{Mirowski et~al.}(1980)\citenamefont{Mirowski, Reid,
  Mower, Watkins, Gott, Schauble, Langer, Heilman, Kolenik, Fischell
  et~al.}}]{defib_nejm}
\bibinfo{author}{\bibfnamefont{M.}~\bibnamefont{Mirowski}},
  \bibinfo{author}{\bibfnamefont{P.}~\bibnamefont{Reid}},
  \bibinfo{author}{\bibfnamefont{M.}~\bibnamefont{Mower}},
  \bibinfo{author}{\bibfnamefont{L.}~\bibnamefont{Watkins}},
  \bibinfo{author}{\bibfnamefont{V.}~\bibnamefont{Gott}},
  \bibinfo{author}{\bibfnamefont{J.}~\bibnamefont{Schauble}},
  \bibinfo{author}{\bibfnamefont{A.}~\bibnamefont{Langer}},
  \bibinfo{author}{\bibfnamefont{M.}~\bibnamefont{Heilman}},
  \bibinfo{author}{\bibfnamefont{S.}~\bibnamefont{Kolenik}},
  \bibinfo{author}{\bibfnamefont{R.}~\bibnamefont{Fischell}},
  \bibnamefont{et~al.}, \bibinfo{journal}{New England Journal of Medicine}
  \textbf{\bibinfo{volume}{303}}, \bibinfo{pages}{322} (\bibinfo{year}{1980}).

\bibitem[{\citenamefont{Thabit et~al.}(2015{\natexlab{a}})\citenamefont{Thabit,
  Tauschman, Allen, Leelarathna, Hartnell, Wilinska, Acerini, Dellweg, Benesch,
  Heinemann et~al.}}]{cc_art_beta}
\bibinfo{author}{\bibfnamefont{H.}~\bibnamefont{Thabit}},
  \bibinfo{author}{\bibfnamefont{M.}~\bibnamefont{Tauschman}},
  \bibinfo{author}{\bibfnamefont{J.}~\bibnamefont{Allen}},
  \bibinfo{author}{\bibfnamefont{L.}~\bibnamefont{Leelarathna}},
  \bibinfo{author}{\bibfnamefont{S.}~\bibnamefont{Hartnell}},
  \bibinfo{author}{\bibfnamefont{M.}~\bibnamefont{Wilinska}},
  \bibinfo{author}{\bibfnamefont{C.}~\bibnamefont{Acerini}},
  \bibinfo{author}{\bibfnamefont{S.}~\bibnamefont{Dellweg}},
  \bibinfo{author}{\bibfnamefont{C.}~\bibnamefont{Benesch}},
  \bibinfo{author}{\bibfnamefont{L.}~\bibnamefont{Heinemann}},
  \bibnamefont{et~al.}, \bibinfo{journal}{New England Journal of Medicine}
  (\bibinfo{year}{2015}{\natexlab{a}}).

\bibitem[{\citenamefont{Leelarathna et~al.}(2013)\citenamefont{Leelarathna,
  English, Thabit, Caldwell, Allen, Kumareswaran, Wilinska, Nodale, Mangat,
  Evans et~al.}}]{closed_loop_glucose_control_ICU}
\bibinfo{author}{\bibfnamefont{L.}~\bibnamefont{Leelarathna}},
  \bibinfo{author}{\bibfnamefont{S.~W.} \bibnamefont{English}},
  \bibinfo{author}{\bibfnamefont{H.}~\bibnamefont{Thabit}},
  \bibinfo{author}{\bibfnamefont{K.}~\bibnamefont{Caldwell}},
  \bibinfo{author}{\bibfnamefont{J.~M.} \bibnamefont{Allen}},
  \bibinfo{author}{\bibfnamefont{K.}~\bibnamefont{Kumareswaran}},
  \bibinfo{author}{\bibfnamefont{M.~E.} \bibnamefont{Wilinska}},
  \bibinfo{author}{\bibfnamefont{M.}~\bibnamefont{Nodale}},
  \bibinfo{author}{\bibfnamefont{J.}~\bibnamefont{Mangat}},
  \bibinfo{author}{\bibfnamefont{M.~L.} \bibnamefont{Evans}},
  \bibnamefont{et~al.}, \bibinfo{journal}{Crit Care}
  \textbf{\bibinfo{volume}{17}}, \bibinfo{pages}{R159} (\bibinfo{year}{2013}).

\bibitem[{\citenamefont{Cobelli et~al.}(2009)\citenamefont{Cobelli, Man,
  Sparacino, Magni, {D}e Nicolao, and Kovatchev}}]{corbelli_review}
\bibinfo{author}{\bibfnamefont{C.}~\bibnamefont{Cobelli}},
  \bibinfo{author}{\bibfnamefont{C.}~\bibnamefont{Man}},
  \bibinfo{author}{\bibfnamefont{G.}~\bibnamefont{Sparacino}},
  \bibinfo{author}{\bibfnamefont{L.}~\bibnamefont{Magni}},
  \bibinfo{author}{\bibfnamefont{G.}~\bibnamefont{{D}e Nicolao}},
  \bibnamefont{and}
  \bibinfo{author}{\bibfnamefont{B.}~\bibnamefont{Kovatchev}},
  \bibinfo{journal}{IEEE Reviews in Biomedical Engineering}
  \textbf{\bibinfo{volume}{2}}, \bibinfo{pages}{54} (\bibinfo{year}{2009}).

\bibitem[{\citenamefont{Thabit et~al.}(2015{\natexlab{b}})\citenamefont{Thabit,
  Tauschman, Allen, Leelarathna, Hartnell, Wilinska, Acerini, Dellweg, Benesch,
  Heinemann et~al.}}]{artificial_beta_cell}
\bibinfo{author}{\bibfnamefont{H.}~\bibnamefont{Thabit}},
  \bibinfo{author}{\bibfnamefont{M.}~\bibnamefont{Tauschman}},
  \bibinfo{author}{\bibfnamefont{J.~M.} \bibnamefont{Allen}},
  \bibinfo{author}{\bibfnamefont{L.}~\bibnamefont{Leelarathna}},
  \bibinfo{author}{\bibfnamefont{S.}~\bibnamefont{Hartnell}},
  \bibinfo{author}{\bibfnamefont{M.~E.} \bibnamefont{Wilinska}},
  \bibinfo{author}{\bibfnamefont{C.~L.} \bibnamefont{Acerini}},
  \bibinfo{author}{\bibfnamefont{S.}~\bibnamefont{Dellweg}},
  \bibinfo{author}{\bibfnamefont{C.}~\bibnamefont{Benesch}},
  \bibinfo{author}{\bibfnamefont{L.}~\bibnamefont{Heinemann}},
  \bibnamefont{et~al.}, \bibinfo{journal}{N. Engl. J. Med.}
  (\bibinfo{year}{2015}{\natexlab{b}}).

\bibitem[{\citenamefont{Glass and Courtemanche}(2000)}]{leon_court}
\bibinfo{author}{\bibfnamefont{L.}~\bibnamefont{Glass}} \bibnamefont{and}
  \bibinfo{author}{\bibfnamefont{M.}~\bibnamefont{Courtemanche}},
  \emph{\bibinfo{title}{Cardiac Arrhythmias and Device Therapy: Results and
  Perspectives for the New Century}} (\bibinfo{publisher}{Futura},
  \bibinfo{year}{2000}), chap. \bibinfo{chapter}{Control of atrial
  fibrillation: A theoretical perspective}, pp. \bibinfo{pages}{87--94}.

\bibitem[{\citenamefont{Christini and Glass}(2002)}]{dave_leon_3}
\bibinfo{author}{\bibfnamefont{D.}~\bibnamefont{Christini}} \bibnamefont{and}
  \bibinfo{author}{\bibfnamefont{L.}~\bibnamefont{Glass}},
  \bibinfo{journal}{Chaos} \textbf{\bibinfo{volume}{12}}, \bibinfo{pages}{732}
  (\bibinfo{year}{2002}).

\bibitem[{\citenamefont{Hall et~al.}(1997)\citenamefont{Hall, Christini,
  Tremblay, Collins, Glass, and Billette}}]{christini_cardio_alternans}
\bibinfo{author}{\bibfnamefont{K.}~\bibnamefont{Hall}},
  \bibinfo{author}{\bibfnamefont{D.~J.} \bibnamefont{Christini}},
  \bibinfo{author}{\bibfnamefont{M.}~\bibnamefont{Tremblay}},
  \bibinfo{author}{\bibfnamefont{J.~J.} \bibnamefont{Collins}},
  \bibinfo{author}{\bibfnamefont{L.}~\bibnamefont{Glass}}, \bibnamefont{and}
  \bibinfo{author}{\bibfnamefont{J.}~\bibnamefont{Billette}},
  \bibinfo{journal}{Phys. Rev. Lett.} \textbf{\bibinfo{volume}{78}},
  \bibinfo{pages}{4938} (\bibinfo{year}{1997}).

\bibitem[{\citenamefont{Mackey and Glass}(1977)}]{mackey_glass_equations}
\bibinfo{author}{\bibfnamefont{M.}~\bibnamefont{Mackey}} \bibnamefont{and}
  \bibinfo{author}{\bibfnamefont{L.}~\bibnamefont{Glass}},
  \bibinfo{journal}{Science} \textbf{\bibinfo{volume}{197}},
  \bibinfo{pages}{287} (\bibinfo{year}{1977}).

\bibitem[{\citenamefont{Parker et~al.}(20xx)\citenamefont{Parker, III, Ward,
  and Peppas}}]{parker_ICU_1}
\bibinfo{author}{\bibfnamefont{R.~S.} \bibnamefont{Parker}},
  \bibinfo{author}{\bibfnamefont{F.~J.~D.} \bibnamefont{III}},
  \bibinfo{author}{\bibfnamefont{J.~F.} \bibnamefont{Ward}}, \bibnamefont{and}
  \bibinfo{author}{\bibfnamefont{N.~A.} \bibnamefont{Peppas}},
  \bibinfo{journal}{AIChE Journa} \textbf{\bibinfo{volume}{46}},
  \bibinfo{pages}{2537} (\bibinfo{year}{20xx}).

\bibitem[{\citenamefont{Parker et~al.}(2001)\citenamefont{Parker, III, , and
  Peppas}}]{parker_doyle}
\bibinfo{author}{\bibfnamefont{R.~S.} \bibnamefont{Parker}},
  \bibinfo{author}{\bibfnamefont{F.~J.~D.} \bibnamefont{III}}, ,
  \bibnamefont{and} \bibinfo{author}{\bibfnamefont{N.~A.}
  \bibnamefont{Peppas}}, \bibinfo{journal}{IEEE Engineering in Medicine and
  Biology} \textbf{\bibinfo{volume}{20}}, \bibinfo{pages}{65}
  (\bibinfo{year}{2001}).

\bibitem[{\citenamefont{Donnet and Samson}(2013)}]{pharmaco_2}
\bibinfo{author}{\bibfnamefont{S.}~\bibnamefont{Donnet}} \bibnamefont{and}
  \bibinfo{author}{\bibfnamefont{A.}~\bibnamefont{Samson}},
  \bibinfo{journal}{Advanced Drug Delivery Reviews}  (\bibinfo{year}{2013}).

\bibitem[{\citenamefont{Bonate}(2005)}]{pharmaco_1}
\bibinfo{author}{\bibfnamefont{P.}~\bibnamefont{Bonate}},
  \bibinfo{journal}{AAPS}  (\bibinfo{year}{2005}).

\bibitem[{\citenamefont{Sadean and Glass}(2009)}]{pharmaco_3}
\bibinfo{author}{\bibfnamefont{M.}~\bibnamefont{Sadean}} \bibnamefont{and}
  \bibinfo{author}{\bibfnamefont{P.}~\bibnamefont{Glass}},
  \bibinfo{journal}{Curr Opin Anaesthesiol}  (\bibinfo{year}{2009}).

\bibitem[{\citenamefont{Kristensen et~al.}(2005)\citenamefont{Kristensen,
  Madsen, , and Ingwersen}}]{pharmaco_4}
\bibinfo{author}{\bibfnamefont{N.~R.} \bibnamefont{Kristensen}},
  \bibinfo{author}{\bibfnamefont{H.}~\bibnamefont{Madsen}}, , \bibnamefont{and}
  \bibinfo{author}{\bibfnamefont{S.~H.} \bibnamefont{Ingwersen}},
  \bibinfo{journal}{XXXX} \textbf{\bibinfo{volume}{32}} (\bibinfo{year}{2005}).

\bibitem[{\citenamefont{Sedigh-Sarvestan
  et~al.}(2012)\citenamefont{Sedigh-Sarvestan, Albers, and
  Gluckman}}]{albers_IEEE}
\bibinfo{author}{\bibfnamefont{M.}~\bibnamefont{Sedigh-Sarvestan}},
  \bibinfo{author}{\bibfnamefont{D.}~\bibnamefont{Albers}}, \bibnamefont{and}
  \bibinfo{author}{\bibfnamefont{B.}~\bibnamefont{Gluckman}}, in
  \emph{\bibinfo{booktitle}{34th Annual International IEEE EMBS Conference}}
  (\bibinfo{year}{2012}).

\bibitem[{\citenamefont{Selgrade et~al.}(2009)\citenamefont{Selgrade, Harris,
  and Pasteur}}]{selgrade_female_endo_data}
\bibinfo{author}{\bibfnamefont{J.~F.} \bibnamefont{Selgrade}},
  \bibinfo{author}{\bibfnamefont{L.~A.} \bibnamefont{Harris}},
  \bibnamefont{and} \bibinfo{author}{\bibfnamefont{R.~D.}
  \bibnamefont{Pasteur}}, \bibinfo{journal}{Journal of Theoretical Biology}
  \textbf{\bibinfo{volume}{260}}, \bibinfo{pages}{572} (\bibinfo{year}{2009}).

\bibitem[{\citenamefont{Sedigh-Sarvestani
  et~al.}(2012)\citenamefont{Sedigh-Sarvestani, Schiff, and
  Gluckman}}]{bruce_sleep}
\bibinfo{author}{\bibfnamefont{M.}~\bibnamefont{Sedigh-Sarvestani}},
  \bibinfo{author}{\bibfnamefont{S.~J.} \bibnamefont{Schiff}},
  \bibnamefont{and} \bibinfo{author}{\bibfnamefont{B.~J.}
  \bibnamefont{Gluckman}}, \bibinfo{journal}{PLoS Comput Biol}
  \textbf{\bibinfo{volume}{8}} (\bibinfo{year}{2012}).

\bibitem[{\citenamefont{Lin et~al.}(2004)\citenamefont{Lin, Chase, Shaw, Doran,
  Hann, Robertson, Browne, Lotz, Wake,  et~al.}}]{chase_icu_2}
\bibinfo{author}{\bibfnamefont{J.}~\bibnamefont{Lin}},
  \bibinfo{author}{\bibfnamefont{J.~G.} \bibnamefont{Chase}},
  \bibinfo{author}{\bibfnamefont{G.~M.} \bibnamefont{Shaw}},
  \bibinfo{author}{\bibfnamefont{C.~V.} \bibnamefont{Doran}},
  \bibinfo{author}{\bibfnamefont{C.~E.} \bibnamefont{Hann}},
  \bibinfo{author}{\bibfnamefont{M.~B.} \bibnamefont{Robertson}},
  \bibinfo{author}{\bibfnamefont{P.~M.} \bibnamefont{Browne}},
  \bibinfo{author}{\bibfnamefont{T.}~\bibnamefont{Lotz}},
  \bibinfo{author}{\bibfnamefont{G.~C.} \bibnamefont{Wake}}, ,
  \bibnamefont{et~al.}, in \emph{\bibinfo{booktitle}{34th Annual International
  IEEE EMBS Conference}} (\bibinfo{year}{2004}).

\bibitem[{\citenamefont{Lin et~al.}(2011)\citenamefont{Lin, Razak, Pretty,
  Compte, Docherty, Parente, Shaw, Hann, , and Chase}}]{chase_icu_1}
\bibinfo{author}{\bibfnamefont{J.}~\bibnamefont{Lin}},
  \bibinfo{author}{\bibfnamefont{N.~N.} \bibnamefont{Razak}},
  \bibinfo{author}{\bibfnamefont{C.~G.} \bibnamefont{Pretty}},
  \bibinfo{author}{\bibfnamefont{A.~L.} \bibnamefont{Compte}},
  \bibinfo{author}{\bibfnamefont{P.}~\bibnamefont{Docherty}},
  \bibinfo{author}{\bibfnamefont{J.~D.} \bibnamefont{Parente}},
  \bibinfo{author}{\bibfnamefont{G.~M.} \bibnamefont{Shaw}},
  \bibinfo{author}{\bibfnamefont{C.~E.} \bibnamefont{Hann}}, ,
  \bibnamefont{and} \bibinfo{author}{\bibfnamefont{G.}~\bibnamefont{Chase}},
  \bibinfo{journal}{Computer Methods and Programs in Biomedicine}
  \textbf{\bibinfo{volume}{102}}, \bibinfo{pages}{192} (\bibinfo{year}{2011}).

\bibitem[{\citenamefont{Levine et~al.}(2017)\citenamefont{Levine, Hripcsak,
  Mamykina, Stuart, and Albers}}]{online_offline_DA}
\bibinfo{author}{\bibfnamefont{M.}~\bibnamefont{Levine}},
  \bibinfo{author}{\bibfnamefont{G.}~\bibnamefont{Hripcsak}},
  \bibinfo{author}{\bibfnamefont{L.}~\bibnamefont{Mamykina}},
  \bibinfo{author}{\bibfnamefont{A.}~\bibnamefont{Stuart}}, \bibnamefont{and}
  \bibinfo{author}{\bibfnamefont{D.}~\bibnamefont{Albers}}
  (\bibinfo{year}{2017}), \bibinfo{note}{arXiv:1709.00163}.

\bibitem[{\citenamefont{Hines et~al.}(2014)\citenamefont{Hines, Middendorf, and
  Aldrich}}]{physio_bayesian_identifiability}
\bibinfo{author}{\bibfnamefont{E.}~\bibnamefont{Hines}},
  \bibinfo{author}{\bibfnamefont{T.}~\bibnamefont{Middendorf}},
  \bibnamefont{and} \bibinfo{author}{\bibfnamefont{R.}~\bibnamefont{Aldrich}},
  \bibinfo{journal}{J Gen. Physiol} \textbf{\bibinfo{volume}{143}},
  \bibinfo{pages}{401} (\bibinfo{year}{2014}).

\bibitem[{\citenamefont{Simens et~al.}(2017)\citenamefont{Simens, Cree-Green,
  Bergman, Nadeau, and Behn}}]{cecelia_ogtt_ident}
\bibinfo{author}{\bibfnamefont{J.}~\bibnamefont{Simens}},
  \bibinfo{author}{\bibfnamefont{M.}~\bibnamefont{Cree-Green}},
  \bibinfo{author}{\bibfnamefont{B.}~\bibnamefont{Bergman}},
  \bibinfo{author}{\bibfnamefont{K.}~\bibnamefont{Nadeau}}, \bibnamefont{and}
  \bibinfo{author}{\bibfnamefont{C.~D.} \bibnamefont{Behn}}, in
  \emph{\bibinfo{booktitle}{Association of Women in Mathematics Annual
  Symposium}} (\bibinfo{year}{2017}).

\bibitem[{\citenamefont{Eisenberg and Hayashi}(2014)}]{marissa_ident}
\bibinfo{author}{\bibfnamefont{M.}~\bibnamefont{Eisenberg}} \bibnamefont{and}
  \bibinfo{author}{\bibfnamefont{M.}~\bibnamefont{Hayashi}},
  \bibinfo{journal}{Math Biosciences} \textbf{\bibinfo{volume}{256}},
  \bibinfo{pages}{116} (\bibinfo{year}{2014}).

\bibitem[{\citenamefont{Walch and Eisenberg}(2016)}]{marissa_2}
\bibinfo{author}{\bibfnamefont{O.}~\bibnamefont{Walch}} \bibnamefont{and}
  \bibinfo{author}{\bibfnamefont{M.}~\bibnamefont{Eisenberg}},
  \bibinfo{journal}{Neurocomputing} \textbf{\bibinfo{volume}{199}},
  \bibinfo{pages}{137} (\bibinfo{year}{2016}).

\bibitem[{\citenamefont{Brouwer et~al.}(2018)\citenamefont{Brouwer, Meza, and
  Eisenberg}}]{marissa_3}
\bibinfo{author}{\bibfnamefont{A.}~\bibnamefont{Brouwer}},
  \bibinfo{author}{\bibfnamefont{R.}~\bibnamefont{Meza}}, \bibnamefont{and}
  \bibinfo{author}{\bibfnamefont{M.}~\bibnamefont{Eisenberg}},
  \bibinfo{journal}{Risk analysis}  (\bibinfo{year}{2018}).

\bibitem[{\citenamefont{Chis et~al.}(2011)\citenamefont{Chis, Banga, and
  Balsa-Canto}}]{structural_id_hard}
\bibinfo{author}{\bibfnamefont{O.-T.} \bibnamefont{Chis}},
  \bibinfo{author}{\bibfnamefont{J.}~\bibnamefont{Banga}}, \bibnamefont{and}
  \bibinfo{author}{\bibfnamefont{E.}~\bibnamefont{Balsa-Canto}},
  \bibinfo{journal}{PLoS ONE} \textbf{\bibinfo{volume}{6}},
  \bibinfo{pages}{e27755} (\bibinfo{year}{2011}).

\bibitem[{\citenamefont{Hamill}(2006)}]{tom_enkf_review}
\bibinfo{author}{\bibfnamefont{T.~M.} \bibnamefont{Hamill}},
  \bibinfo{journal}{XXXX}  (\bibinfo{year}{2006}).

\bibitem[{\citenamefont{Kuznetzov}(1998)}]{kuzbook}
\bibinfo{author}{\bibfnamefont{Y.}~\bibnamefont{Kuznetzov}},
  \emph{\bibinfo{title}{Bifurcation Theory}}
  (\bibinfo{publisher}{Springer-Verlag}, \bibinfo{year}{1998}),
  \bibinfo{edition}{$2^{nd}$} ed.

\bibitem[{\citenamefont{Hastie et~al.}(2015)\citenamefont{Hastie, Tibshirani,
  and Wainwright}}]{statistical_learning_sparsity}
\bibinfo{author}{\bibfnamefont{T.}~\bibnamefont{Hastie}},
  \bibinfo{author}{\bibfnamefont{R.}~\bibnamefont{Tibshirani}},
  \bibnamefont{and}
  \bibinfo{author}{\bibfnamefont{M.}~\bibnamefont{Wainwright}},
  \emph{\bibinfo{title}{Statistical learning with sparsity}}
  (\bibinfo{publisher}{CRC}, \bibinfo{year}{2015}).

\bibitem[{\citenamefont{Mamykina et~al.}(2015)\citenamefont{Mamykina, Levine,
  Davidson, Smaldone, Elhadad, and Albers}}]{me_lena_1}
\bibinfo{author}{\bibfnamefont{L.}~\bibnamefont{Mamykina}},
  \bibinfo{author}{\bibfnamefont{M.}~\bibnamefont{Levine}},
  \bibinfo{author}{\bibfnamefont{P.}~\bibnamefont{Davidson}},
  \bibinfo{author}{\bibfnamefont{A.}~\bibnamefont{Smaldone}},
  \bibinfo{author}{\bibfnamefont{N.}~\bibnamefont{Elhadad}}, \bibnamefont{and}
  \bibinfo{author}{\bibfnamefont{D.}~\bibnamefont{Albers}}, \bibinfo{journal}{J
  Am Med Inform Asso}  (\bibinfo{year}{2015}).

\bibitem[{\citenamefont{Young}(2002)}]{youngSRB}
\bibinfo{author}{\bibfnamefont{L.-S.} \bibnamefont{Young}},
  \bibinfo{journal}{J. Stat. Phys.} \textbf{\bibinfo{volume}{108}},
  \bibinfo{pages}{733} (\bibinfo{year}{2002}).

\bibitem[{\citenamefont{Ruelle}(1976)}]{ruelle_srb1}
\bibinfo{author}{\bibfnamefont{D.}~\bibnamefont{Ruelle}},
  \bibinfo{journal}{Amer. {J}. {M}ath} \textbf{\bibinfo{volume}{98}},
  \bibinfo{pages}{619} (\bibinfo{year}{1976}).

\bibitem[{\citenamefont{Sinai}(1972)}]{sinai_srb_1}
\bibinfo{author}{\bibfnamefont{Y.}~\bibnamefont{Sinai}},
  \bibinfo{journal}{Russian Math. Surveys} \textbf{\bibinfo{volume}{27}},
  \bibinfo{pages}{21} (\bibinfo{year}{1972}).

\bibitem[{\citenamefont{Bowen}(1975)}]{bowen_srb1}
\bibinfo{author}{\bibfnamefont{R.}~\bibnamefont{Bowen}},
  \emph{\bibinfo{title}{Equilibrium states and teh ergodic theory of {A}nosov
  diffeomorphisms}}, vol. \bibinfo{volume}{470} of \emph{\bibinfo{series}{Lect.
  Notes in Math.}} (\bibinfo{publisher}{Springer-{V}erlag},
  \bibinfo{address}{Berlin}, \bibinfo{year}{1975}).

\bibitem[{\citenamefont{Sturis et~al.}(1991)\citenamefont{Sturis, Polonsky,
  Mosekilde, and Cauter}}]{sturis_91}
\bibinfo{author}{\bibfnamefont{J.}~\bibnamefont{Sturis}},
  \bibinfo{author}{\bibfnamefont{K.~S.} \bibnamefont{Polonsky}},
  \bibinfo{author}{\bibfnamefont{E.}~\bibnamefont{Mosekilde}},
  \bibnamefont{and} \bibinfo{author}{\bibfnamefont{E.~V.}
  \bibnamefont{Cauter}}, \bibinfo{journal}{Am J Physiol Endocrinol Metab}
  \textbf{\bibinfo{volume}{260}}, \bibinfo{pages}{E801} (\bibinfo{year}{1991}).

\bibitem[{\citenamefont{Julier and Uhlmann}(2004)}]{ukf_review}
\bibinfo{author}{\bibfnamefont{S.}~\bibnamefont{Julier}} \bibnamefont{and}
  \bibinfo{author}{\bibfnamefont{J.}~\bibnamefont{Uhlmann}},
  \bibinfo{journal}{Proc. IEEE} \textbf{\bibinfo{volume}{92}},
  \bibinfo{pages}{401} (\bibinfo{year}{2004}).

\bibitem[{\citenamefont{Julier et~al.}(1995)\citenamefont{Julier, Uhlmann, and
  Durrant-Whyte}}]{ukf_o}
\bibinfo{author}{\bibfnamefont{S.}~\bibnamefont{Julier}},
  \bibinfo{author}{\bibfnamefont{J.}~\bibnamefont{Uhlmann}}, \bibnamefont{and}
  \bibinfo{author}{\bibfnamefont{H.}~\bibnamefont{Durrant-Whyte}}, in
  \emph{\bibinfo{booktitle}{American Control Conference}},
  \bibinfo{organization}{IEEE} (\bibinfo{publisher}{IEEE},
  \bibinfo{year}{1995}).

\bibitem[{\citenamefont{Want and Merwe}(2000)}]{wan_primary_ukf}
\bibinfo{author}{\bibfnamefont{E.}~\bibnamefont{Want}} \bibnamefont{and}
  \bibinfo{author}{\bibfnamefont{R.}~\bibnamefont{Merwe}}, in
  \emph{\bibinfo{booktitle}{Adaptive Systems for Signal Processing,
  Communications, and Control Symposium}}, \bibinfo{organization}{IEEE}
  (\bibinfo{publisher}{Wiley}, \bibinfo{year}{2000}), pp.
  \bibinfo{pages}{153--158}.

\bibitem[{\citenamefont{Gove and Hollinger}(2006)}]{dual_ukf}
\bibinfo{author}{\bibfnamefont{J.}~\bibnamefont{Gove}} \bibnamefont{and}
  \bibinfo{author}{\bibfnamefont{D.}~\bibnamefont{Hollinger}},
  \bibinfo{journal}{J Geophys Res} \textbf{\bibinfo{volume}{111}},
  \bibinfo{pages}{DO8S07} (\bibinfo{year}{2006}).

\bibitem[{\citenamefont{Want et~al.}(2000)\citenamefont{Want, Merwe, and
  Nelson}}]{wan_dual_ukf_wins}
\bibinfo{author}{\bibfnamefont{E.}~\bibnamefont{Want}},
  \bibinfo{author}{\bibfnamefont{R.}~\bibnamefont{Merwe}}, \bibnamefont{and}
  \bibinfo{author}{\bibfnamefont{A.}~\bibnamefont{Nelson}}, in
  \emph{\bibinfo{booktitle}{NEURAL INFORMATION PROCESSING SYSTEMS}}
  (\bibinfo{publisher}{MIT Press}, \bibinfo{year}{2000}), pp.
  \bibinfo{pages}{666--672}.

\bibitem[{\citenamefont{Wan and Merwe}(2001)}]{kalman_nn_ukf}
\bibinfo{author}{\bibfnamefont{E.~A.} \bibnamefont{Wan}} \bibnamefont{and}
  \bibinfo{author}{\bibfnamefont{R.~V.~D.} \bibnamefont{Merwe}}, in
  \emph{\bibinfo{booktitle}{Kalman Filtering and Neural Networks}}
  (\bibinfo{publisher}{Wiley}, \bibinfo{year}{2001}), pp.
  \bibinfo{pages}{221--280}.

\bibitem[{\citenamefont{Cotter et~al.}(2013)\citenamefont{Cotter, Roberts,
  Stuart, and White}}]{cotter_mcmc_faster}
\bibinfo{author}{\bibfnamefont{S.~L.} \bibnamefont{Cotter}},
  \bibinfo{author}{\bibfnamefont{G.~O.} \bibnamefont{Roberts}},
  \bibinfo{author}{\bibfnamefont{A.~M.} \bibnamefont{Stuart}},
  \bibnamefont{and} \bibinfo{author}{\bibfnamefont{D.}~\bibnamefont{White}},
  \bibinfo{journal}{Statist. Sci.} \textbf{\bibinfo{volume}{28}},
  \bibinfo{pages}{424} (\bibinfo{year}{2013}).

\bibitem[{\citenamefont{Sotomayor}(1973)}]{soyomayorbifsets}
\bibinfo{author}{\bibfnamefont{J.}~\bibnamefont{Sotomayor}}, in
  \emph{\bibinfo{booktitle}{Dynamical Systems}} (\bibinfo{year}{1973}), pp.
  \bibinfo{pages}{549--560}.

\bibitem[{\citenamefont{Pugh and Shub}(1989)}]{p-s-ergodic-attractors}
\bibinfo{author}{\bibfnamefont{C.}~\bibnamefont{Pugh}} \bibnamefont{and}
  \bibinfo{author}{\bibfnamefont{M.}~\bibnamefont{Shub}},
  \bibinfo{journal}{Trans. Amer. Math. Soc.} \textbf{\bibinfo{volume}{312}},
  \bibinfo{pages}{1} (\bibinfo{year}{1989}).

\bibitem[{\citenamefont{Burns and Wilkinson}(2010)}]{burns_wilk_annals}
\bibinfo{author}{\bibfnamefont{K.}~\bibnamefont{Burns}} \bibnamefont{and}
  \bibinfo{author}{\bibfnamefont{A.}~\bibnamefont{Wilkinson}},
  \bibinfo{journal}{Annals of Mathematics}  (\bibinfo{year}{2010}).

\bibitem[{\citenamefont{Albers and Sprott}(2006)}]{hypviolation}
\bibinfo{author}{\bibfnamefont{D.~J.} \bibnamefont{Albers}} \bibnamefont{and}
  \bibinfo{author}{\bibfnamefont{J.~C.} \bibnamefont{Sprott}},
  \bibinfo{journal}{Nonlinearity} \textbf{\bibinfo{volume}{19}},
  \bibinfo{pages}{1801} (\bibinfo{year}{2006}).

\bibitem[{\citenamefont{Albers et~al.}(2006)\citenamefont{Albers, Sprott, and
  Crutchfield}}]{dynamicsPRL}
\bibinfo{author}{\bibfnamefont{D.~J.} \bibnamefont{Albers}},
  \bibinfo{author}{\bibfnamefont{J.~C.} \bibnamefont{Sprott}},
  \bibnamefont{and} \bibinfo{author}{\bibfnamefont{J.~P.}
  \bibnamefont{Crutchfield}}, \bibinfo{journal}{Phys. Rev. E}
  \textbf{\bibinfo{volume}{74}}, \bibinfo{pages}{057201}
  (\bibinfo{year}{2006}).

\bibitem[{\citenamefont{Tibshirani}(1996)}]{lasso_1}
\bibinfo{author}{\bibfnamefont{R.}~\bibnamefont{Tibshirani}},
  \bibinfo{journal}{Journal of the Royal Statistical Society, Series B}
  \textbf{\bibinfo{volume}{58}}, \bibinfo{pages}{267} (\bibinfo{year}{1996}).

\bibitem[{\citenamefont{Breiman}(1995)}]{lasso_2}
\bibinfo{author}{\bibfnamefont{L.}~\bibnamefont{Breiman}},
  \bibinfo{journal}{Technometrics} \textbf{\bibinfo{volume}{37}},
  \bibinfo{pages}{373} (\bibinfo{year}{1995}).

\bibitem[{\citenamefont{Zou and Hastie}(2005)}]{elastic_nets}
\bibinfo{author}{\bibfnamefont{H.}~\bibnamefont{Zou}} \bibnamefont{and}
  \bibinfo{author}{\bibfnamefont{T.}~\bibnamefont{Hastie}},
  \bibinfo{journal}{Journal of the Royal Statistical Society, Series B} pp.
  \bibinfo{pages}{301--320} (\bibinfo{year}{2005}).

\bibitem[{\citenamefont{Jaggi}(2014)}]{elastic_nets_are_linear_svm_1}
\bibinfo{author}{\bibfnamefont{M.}~\bibnamefont{Jaggi}},
  \emph{\bibinfo{title}{An Equivalence between the Lasso and Support Vector
  Machines.}} (\bibinfo{publisher}{Chapman and Hall/CRC},
  \bibinfo{year}{2014}).

\bibitem[{\citenamefont{Pearson}(1901)}]{pca_original}
\bibinfo{author}{\bibfnamefont{K.}~\bibnamefont{Pearson}},
  \bibinfo{journal}{Philosophical Magazine} \textbf{\bibinfo{volume}{2}},
  \bibinfo{pages}{559} (\bibinfo{year}{1901}).

\bibitem[{\citenamefont{Hotelling}(1935)}]{pca_original_2}
\bibinfo{author}{\bibfnamefont{H.}~\bibnamefont{Hotelling}},
  \bibinfo{journal}{J. Ed. Psych.}  (\bibinfo{year}{1935}).

\bibitem[{\citenamefont{Jolliffe}(2010)}]{pca_book}
\bibinfo{author}{\bibfnamefont{I.}~\bibnamefont{Jolliffe}},
  \emph{\bibinfo{title}{Principal Component Analysis}}
  (\bibinfo{publisher}{Springer}, \bibinfo{year}{2010}).

\bibitem[{\citenamefont{Hoeting et~al.}(1999)\citenamefont{Hoeting, Madigan,
  Raftery, and Volinsky}}]{bayesian_model_average_mad}
\bibinfo{author}{\bibfnamefont{J.}~\bibnamefont{Hoeting}},
  \bibinfo{author}{\bibfnamefont{D.}~\bibnamefont{Madigan}},
  \bibinfo{author}{\bibfnamefont{A.}~\bibnamefont{Raftery}}, \bibnamefont{and}
  \bibinfo{author}{\bibfnamefont{C.}~\bibnamefont{Volinsky}},
  \bibinfo{journal}{Statistical Science} \textbf{\bibinfo{volume}{14}},
  \bibinfo{pages}{382} (\bibinfo{year}{1999}).

\bibitem[{\citenamefont{Raftery et~al.}(2005)\citenamefont{Raftery, Gneiting,
  Balabdaoui, and Polakowski}}]{model_average_calibrate_aos}
\bibinfo{author}{\bibfnamefont{A.}~\bibnamefont{Raftery}},
  \bibinfo{author}{\bibfnamefont{T.}~\bibnamefont{Gneiting}},
  \bibinfo{author}{\bibfnamefont{F.}~\bibnamefont{Balabdaoui}},
  \bibnamefont{and}
  \bibinfo{author}{\bibfnamefont{M.}~\bibnamefont{Polakowski}},
  \bibinfo{journal}{Monthly Weather Review} \textbf{\bibinfo{volume}{133}},
  \bibinfo{pages}{1155} (\bibinfo{year}{2005}).

\bibitem[{\citenamefont{Claeskens and Hjort}(2008)}]{claeskens_model_selection}
\bibinfo{author}{\bibfnamefont{G.}~\bibnamefont{Claeskens}} \bibnamefont{and}
  \bibinfo{author}{\bibfnamefont{N.}~\bibnamefont{Hjort}},
  \emph{\bibinfo{title}{Model selection and model averaging}}
  (\bibinfo{publisher}{Cambridge Univesity Press}, \bibinfo{year}{2008}).

\end{thebibliography}






% \end{thebibliography}

\end{document}